\documentclass{emulateapj}
\usepackage{calc}
\newcommand{\dm}{$\Delta m_{15}(B)$}
\newcommand{\ew}{$pW$}
\newcommand{\sneia}{SNe~Ia}
\newcommand{\omm}{$\Omega_M$}
\newcommand{\oml}{$\Omega_\Lambda$}

\shorttitle{Unburned Material in SNe~Ia}
\shortauthors{Folatelli et al.}

\begin{document}

\title{Unburned Material in the Ejecta of Type Ia Supernovae\altaffilmark{1}}

\author{%
  Gast\'on Folatelli\altaffilmark{2,3},
  M. M. Phillips\altaffilmark{4},
  Nidia Morrell\altaffilmark{4},
  Masaomi Tanaka\altaffilmark{2},
  Keiichi Maeda\altaffilmark{2},
  Ken'ichi Nomoto\altaffilmark{2},
  Maximilian Stritzinger\altaffilmark{5,6},
  Christopher R. Burns\altaffilmark{7},
  Mario Hamuy\altaffilmark{3},
  Paolo Mazzali\altaffilmark{8,9}
  Luis Boldt\altaffilmark{10},
  Abdo Campillay\altaffilmark{4},
  Carlos Contreras\altaffilmark{11},
  Sergio Gonz\'alez\altaffilmark{4},
  Miguel Roth\altaffilmark{4},
  Francisco Salgado\altaffilmark{12},
  W. L. Freedman\altaffilmark{7},
  Barry F Madore\altaffilmark{7,13},
  S. E. Persson\altaffilmark{7},
  and 
  Nicholas B. Suntzeff\altaffilmark{14},
}
\altaffiltext{1}{This paper includes data gathered with the 6.5-m
  Magellan Telescopes located at Las Campanas Observatory, Chile; and
  the Gemini Observatory, Cerro Pachon, Chile (Gemini Program
  GS-2008B$-$Q$-$56). Based on observations collected at the European
  Organisation for Astronomical Research in the Southern Hemisphere,
  Chile (ESO Programmes 076.A-0156, 078.D-0048, 080.A-0516, and
  082.A-0526)}
\altaffiltext{2}{Institute for the Physics and Mathematics of the
  Universe (IPMU), University of Tokyo, 5-1-5 Kashiwanoha, Kashiwa,
  Chiba, 277-8583, Japan}
\altaffiltext{3}{Departamento de Astronom\'{\i}a, Universidad de Chile, 
  Casilla 36-D, Santiago, Chile }
\altaffiltext{4}{Las Campanas Observatory, Carnegie Observatories,
  Casilla 601, La Serena, Chile}
\altaffiltext{5}{The Oskar Klein Centre, Department of Astronomy,
  Stockholm University, AlbaNova, 10691, Stockholm, Sweden}
\altaffiltext{6}{Dark Cosmology Centre, Niels Bohr Institute, University
of Copenhagen, Juliane Maries Vej 30, 2100 Copenhagen \O, Denmark.}
\altaffiltext{7}{Observatories of the Carnegie Institution of
  Washington, 813 Santa Barbara St., Pasadena, CA 91101.}
\altaffiltext{8}{Max-Planck Institut f\"ur Astrophysik,
  Karl-Schwarzschild-Str. 1, 85748 Garching, Germany}
\altaffiltext{9}{Istituto Naz. di Astrofisica-Oss. Astron., vicolo
  dell'Osservatorio, 5, 35122 Padova, Italy}
\altaffiltext{10}{Argelander Institut f\"ur Astronomie, Universit\"at
  Bonn, Auf dem H\"ugel 71, D-53111 Bonn, Germany} 
\altaffiltext{11}{Centre for Astrophysics \& Supercomputing, Swinburne
  University of Technology, P.O. Box 218, Victoria 3122, Australia}
\altaffiltext{12}{Leiden Observatory, Leiden University, PO Box 9513,
  NL-2300 RA Leiden, The Netherlands}
\altaffiltext{13}{Infrared Processing and Analysis Center, Caltech/Jet
  Propulsion Laboratory, Pasadena, CA 91125.}
\altaffiltext{14}{George P. and Cynthia Woods Mitchell Institute for
  Fundamental Physics and Astronomy, Department of Physics and
  Astronomy, Texas A\&M University, College Station, TX 77843, USA}

\email{gaston.folatelli@ipmu.jp}

\setcounter{footnote}{15}

\begin{abstract}
\noindent The presence of unburned material in the ejecta of 
  normal Type Ia supernovae (\sneia) is investigated using early-time
  spectroscopy obtained by the {\em Carnegie Supernova Project} (CSP).
  The tell-tale signature of pristine material from a C+O white
    dwarf progenitor star is the presence of carbon, as oxygen is also
    a product of carbon burning. The most prominent carbon lines in
    optical spectra of \sneia\ are expected to arise from \ion{C}{2}.
  We find that at least 30\% of the objects in the 
  sample show an absorption at $\approx$6300 \AA\ which is
    attributed to \ion{C}{2} $\lambda$6580. An alternative
  identification of this absorption as H$\alpha$ is considered to be
  unlikely. These findings imply a larger incidence of carbon in
  \sneia\ ejecta than previously noted. We show how
    observational biases and physical conditions 
  may hide the presence of weak \ion{C}{2} lines, and account for the
  scarcity of previous carbon detections in the literature.
This relatively large frequency of carbon detections has crucial
implications on our understanding of the explosive process. 
Furthermore, the identification of the 6300 \AA\ absorptions as carbon
  would imply that unburned material is present at very low expansion
velocities, merely $\approx$1000 km s$^{-1}$ above the bulk of
\ion{Si}{2}. Based on spectral modeling, it is found that the
detections are consistent with a mass of carbon of
$10^{-3}$ -- $10^{-2} \, M_\odot$. The presence of this material
so deep in the ejecta would imply substantial mixing, which may be
related to asymmetries of the flame propagation. Another possible
explanation for the carbon absorptions may be 
the existence of clumps of unburned material along the line of
sight. However, the uniformity of the relation between \ion{C}{2} and
\ion{Si}{2} velocities is not consistent with such small-scale
  asymmetries. The spectroscopic and photometric properties of
  \sneia\ with and without carbon signatures are compared. A trend
toward bluer color and lower luminosity at maximum light is found for
objects which show carbon.  
\end{abstract}


\section{INTRODUCTION}
\label{sec:intro}

The generally favored picture of a Type Ia SN (SN~Ia) is the
thermonuclear explosion of a C+O white dwarf (WD) star as 
proposed by \citet{hoyle60}. The WD is assumed to belong to a binary
system, and to explode when it reaches the Chandrasekhar mass limit
($M_{\mathrm{Ch}}$) via two main, alternative channels. In the
single-degenerate scenario, a main-sequence or giant companion star
transfers mass to the WD by Roche-Lobe overflow
\citep{whelan73,nomoto82,iben84}. Alternatively,
in the double-degenerate scenario, the companion star may be another
WD, and the $M_{\mathrm{Ch}}$ limit is reached or exceeded when both
objects coalesce \citep{iben84,webbink84}. The actual incidence of each of these
evolutionary paths is currently under debate.

Further controversy regarding the physical nature of \sneia\ is
related to the mechanism by which the thermonuclear flame propagates
inside the WD \citep[see][and references therein]{hillebrandt00}. It
is generally believed the explosion initiates as a 
deflagration, or subsonic flame \citep{nomoto76}. If the WD
  explodes as a prompt detonation, then no intermediate-mass elements
would be left in the outer ejecta, as opposed to the observational
evidence, and in contradiction with the 
chemical evolution of the interstellar medium \citep{arnett71}. A pure
deflagration, however, would not provide the necessary energy and
nucleosynthesis to explain most normal \sneia\ \citep[see,
    e.g.][]{gamezo05}. It seems that a delayed detonation,
i.e. a transition to a supersonic regime after the WD is pre-expanded in
the deflagration phase, is necessary in order to reproduce the 
typical and bright \sneia\ \citep{khokhlov91,hoeflich95}. The mechanism
that produces the transition is not completely understood, mostly due
to the complexity of flame-propagation physics and the limited
resolution of current models. 

Despite these unknowns, \sneia\ are successfully employed as precision
cosmological distance indicators due to their homogeneous
properties. Their peak luminosities can be calibrated through the
observed light-curve width and color to provide distances with a
precision of $\sim$$0.1$ mag or better \citep[e.g., see][and references
  therein]{folatelli10b}. This calibrated precision led to the 
discovery of the acceleration in the current expansion rate of the
Universe \citep{riess98,perlmutter99}.

Some spectral characteristics, such as the strength of \ion{Si}{2}
lines near maximum light, show variations that are correlated with
photometric properties. Temperature in the ejecta has
been identified as a possible driver of this
luminosity-related spectral diversity \citep{nugent95}. Temperature in turn
depends on the amount of $^{56}$Ni synthesized during the
explosion, thus providing a neat physical picture for \sneia. However,
this one-parameter picture is insufficient to 
describe inhomogeneities in other photometric and spectroscopic
properties that have been observed to vary independently of
luminosity.  Moreover, the connection of $^{56}$Ni mass with
  temperature produces changes in the opacity which lead to more
  complex effects in the observed properties
  \citep{hoeflich96,pinto00}. \citet{kasen07} performed a detailed
  study of the line blanketing effects on the light curves and spectra. 

The origin of these inhomogeneities has recently been linked to varying
properties of the progenitor and explosion mechanism 
\citep[e.g., see][]{hoeflich10,maeda11}. The latter determines the
extent of burning and the degree of mixing of ashes and unburned fuel
within the ejecta. During the last decade, multi-dimensional
calculations have been introduced in an attempt to fully reproduce the 
burning process from first principles \citep[see][and references
  therein]{maeda10b}. These models, although 
still relatively rudimentary, naturally introduce a variety of effects
which may succeed in reproducing several aspects of 
the observed spectral diversity. For example, an asymmetric explosion, 
in which the thermonuclear reaction is ignited at an offset from the
WD center, provides an explanation not only for the observed scatter
of the peak magnitudes \citep{kasen09}, but also for the diverse 
velocity and velocity gradient seen in the \ion{Si}{2} $\lambda$6355
line \citep{maeda10c}.

It is an observational fact that spectral inhomogeneities are 
stronger before or near maximum light than later on
\citep{benetti05,branch07}. At these early stages the outer part of
the ejecta is observed. Under the assumption of homologous expansion
\citep[e.g., see][]{roepke05}, this outer material moves at high velocities
($v \gtrsim $10000 km s$^{-1}$). As a consequence, observations
indicate that, whatever the variations in the progenitor system and
explosion mechanism are, the most noticeable effects appear in the
fast-expanding, outer regions of the ejecta.
Variations are commonly observed among
normal \sneia\ in the relative strength of spectral lines, the evolution of
expansion velocities, the appearance of high-velocity components, and
the presence of different ionization states, among other properties
\citep[e.g., see][]{nugent95,benetti05,mazzali05}. The study of pre-maximum
spectra thus promises to be very useful in the validation of
proposed models.  

A relevant characteristic which can be studied using early-time
spectra is the possible presence of unburned material in the outer
ejecta. Most of the available models predict that part of the primordial WD
material composed of carbon and oxygen is left unburned. The amount and
location of this unburned material has direct implications in
the validity of the models. Since oxygen is also
a product of carbon burning, it is the detection of carbon what would most
clearly evidence the existence of unburned material. Furthermore,
singly-ionized states of both C and O are expected to be dominant in
the outer ejecta \citep[$v>$10000 km s$^{-1}$;][]{tanaka08}. Due to
high excitation temperatures of \ion{O}{2}, however, no strong lines
of this ion appear in the optical range. Some strong \ion{O}{1}
transitions produce observed absorptions
which are blended with lines of \ion{Mg}{2}, another product of carbon
burning. At sufficiently low temperature, \ion{C}{1} lines are expected
to appear, most prominently in the near-IR \citep{marion06}, however this
condition is not likely to occur in the regions of the ejecta probed
before maximum light. The same is true for the high-temperature
condition required to produce \ion{C}{3} lines, unless non-LTE effects
play an important role. Consequently, the possible detection of unburned
material must rely on the presence of \ion{C}{2} lines. 

Until recently, the detection of carbon lines in
SN-Ia spectra was uncommon. \ion{C}{2} lines have been clearly 
identified in optical spectra of a few peculiarly 
luminous, slowly expanding objects which have been suggested to arise from
super-Chandrasekhar mass progenitors
\citep[e.g.,][]{howell06,hicken07,yamanaka09b,scalzo10}. In normal 
objects, the evidence of carbon is given by a weak absorption feature
at $\approx$6300 \AA\, attributed to \ion{C}{2} $\lambda$6580
\citep{mazzali01,branch03,thomas07}. An alternative
  identification of this absorption as H$\alpha$ was suggested by
  \citet{garavini04} for SN~1999aa at 11 days before maximum
  light. The evidence, however, was not conclusive. Hydrogen from the
  companion star or a circumstellar envelope could be mixed in the
  outer layers of the ejecta and possibly be detected in spectra at
  phases between 15 and 5 days before maximum. Based on spectral
  synthesis calculations, \citet{lentz02} 
  showed that H$\alpha$ can produce the 6300 \AA\ absorption at such
  early phases if they assumed the existence of $\sim$10$^{-2}$
  $M_\odot$ of hydrogen at $v>15000$ km s$^{-1}$. However, such a
  large amount of hydrogen is in contradiction with the limit of
  $\sim$10$^{-7}$ $M_\odot$ imposed by mass-transfer models
  \citep{nomoto07}. 

In this paper, we analyze the presence of the \ion{C}{2}
  $\lambda$6580 line in a sample of pre-maximum spectra
obtained by the {\em Carnegie Supernova Project}
\citep[CSP;][]{hamuy06}, and thereby infer 
the incidence of unburned material in normal \sneia. In
Section~\ref{sec:data} we make a brief description of the spectroscopic
data, most of which will be presented by \citet{folatelli11}, and
provide the evidence supporting the detection of \ion{C}{2}
lines. The alternative of hydrogen is also studied. In
Section~\ref{sec:spprop} we analyze the spectroscopic properties of
\sneia\ with carbon in comparison with those which show no clear
evidence of it. Section~\ref{sec:phprop} presents a comparative study
of photometric properties of SNe with and without carbon. Finally, in
Section~\ref{sec:dis} we briefly discuss our results in light of the
available explosion models. 
Note that between the presentation of a preliminary report on this work 
\citep{folatelli10a} and submission of the current paper,
\citet{parrent11} published similar results to the ones presented here,
confirming a significantly higher incidence of carbon in \sneia\ spectra 
than previously noted.  In Section~\ref{sec:dis}, we compare
the findings of the paper by \citet{parrent11} with the conclusions
of our own study.

\section{DATA SET AND CARBON DETECTION}
\label{sec:data}

We base the current work on the sample of optical spectra presented by
\citet{folatelli11}. This sample consists of 588 spectra of
82 \sneia\ observed by the CSP between 2004 and 2009. In searching
for signatures of unburned material, we only consider the
pre-maximum phases. This is because any primordial C-O rich
material left by the explosion should lie in the outer parts of the
ejecta, and is thus most easily detectable at the earliest stages
\citep{tanaka08}. For completeness, we have added to the sample of
\citet{folatelli11} other spectra in the CSP database which were
obtained before maximum light. These additional spectra belong to
SNe~2008hj, 2009I, and 2009cz. The present sample amounts to 157
spectra of 51 \sneia\ with ages between $-12$ and $0$
days. Table~\ref{tab:sne} provides the list of SNe used in this
  work. The same table also gives the epoch of the first spectrum, the
class of carbon detection as defined in
Section~\ref{sec:CIIse}, and several photometric parameters which are
used in Section~\ref{sec:phprop}. Table~\ref{tab:spec} contains
information of each spectrum.

We have also included SN~2009dc which is a luminous, slowly expanding
SN~Ia suggested to arise from a super-Chandrasekhar mass progenitor
\citep{yamanaka09b,tanaka10,silverman11,taubenberger11}. For
consistency, we only include the pre-maximum spectra of this SN in
Table~\ref{tab:spec}, although it showed strong \ion{C}{2} features
which persisted after maximum light. In light of the peculiar
nature of this object, we will treat it in Section~\ref{sec:spprop}
as a reference case to see whether there is a connection with the
normal \sneia\ studied here. 

\subsection{The Search for \ion{C}{2}}
\label{sec:CIIse}

As mentioned in Section~\ref{sec:intro}, carbon is the only
  element which can be identified with purely unburned material from
  a C+O WD progenitor. Oxygen instead could be due to a mixture of
  pristine material and ashes from C burning. According to
  \citet{tanaka08}, the dominant 
  carbon ionization state in the outer parts of the ejecta is
  \ion{C}{2}. Therefore, in our study of pre-maximum spectra, we
  search for signatures of unburned material in the form of \ion{C}{2}
  features. 

The four strongest \ion{C}{2} lines at optical wavelengths are: \ion{C}{2}
$\lambda$4267, $\lambda$4745, $\lambda$6580, and $\lambda$7234. 
  Of these, the most prominent in \sneia\ spectra is expected to be \ion{C}{2} 
$\lambda$6580 \citep[e.g., see][]{hatano99,mazzali01}. Even if \ion{C}{2} 
$\lambda$4745 has a lower oscillator strength than \ion{C}{2}
  $\lambda$6580, it may become comparably strong at temperatures of
  $\approx$5000 K due to its lower excitation energy. However, at
  such low temperatures, the blue 
  part of the spectrum is dominated by \ion{Ti}{2} lines, which
  complicates the identification of this carbon feature.
For expansion velocities between 10000 and 20000 km s$^{-1}$ the
\ion{C}{2} $\lambda$6580 line should produce absorption 
between 6170 \AA\ and 6370 \AA, which lies on the red side of the
\ion{Si}{2} $\lambda$6355 line. We have therefore focused the search
for unburned material in this wavelength region.

\begin{figure}[htpb]
\epsscale{1.0}
\plottwo{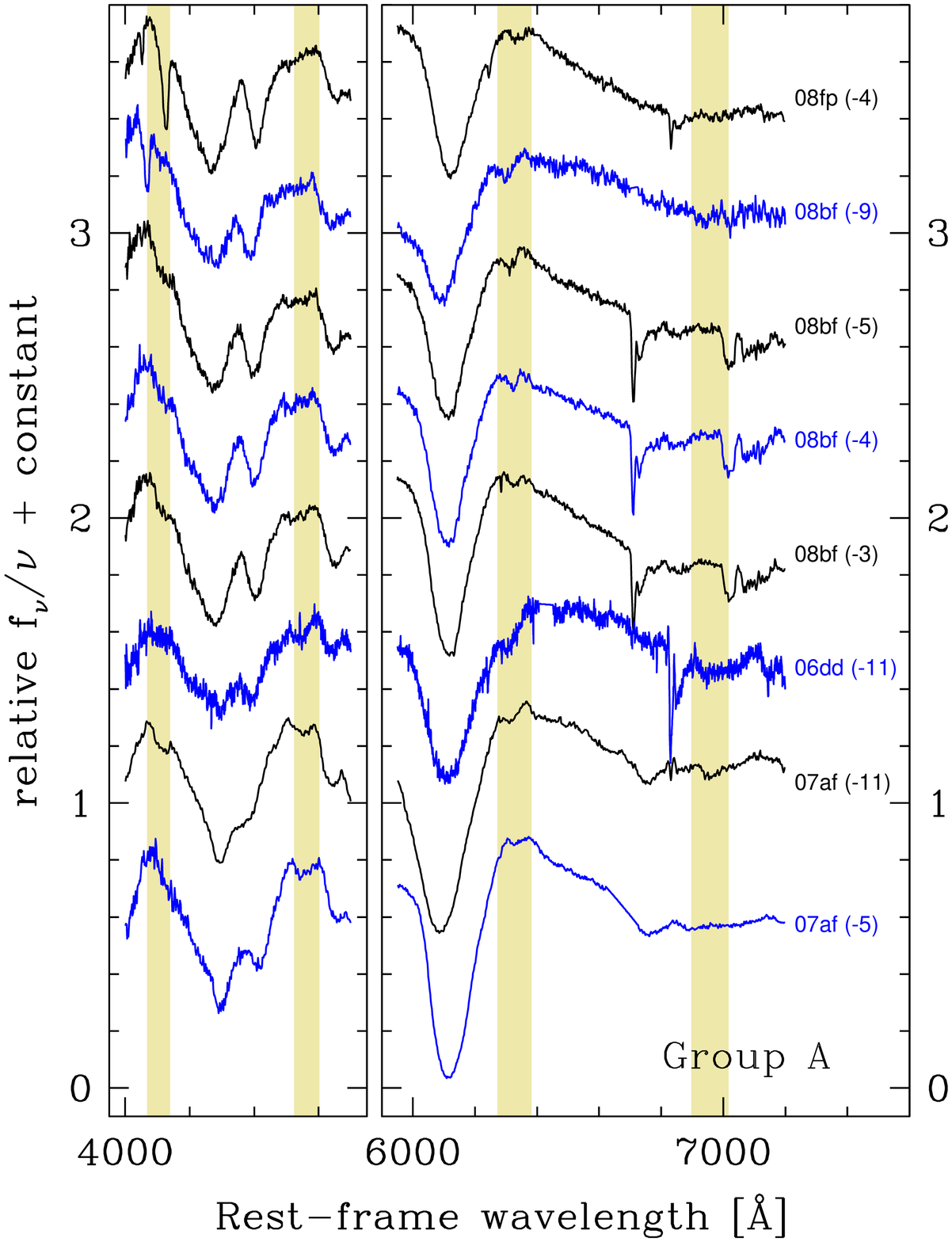}{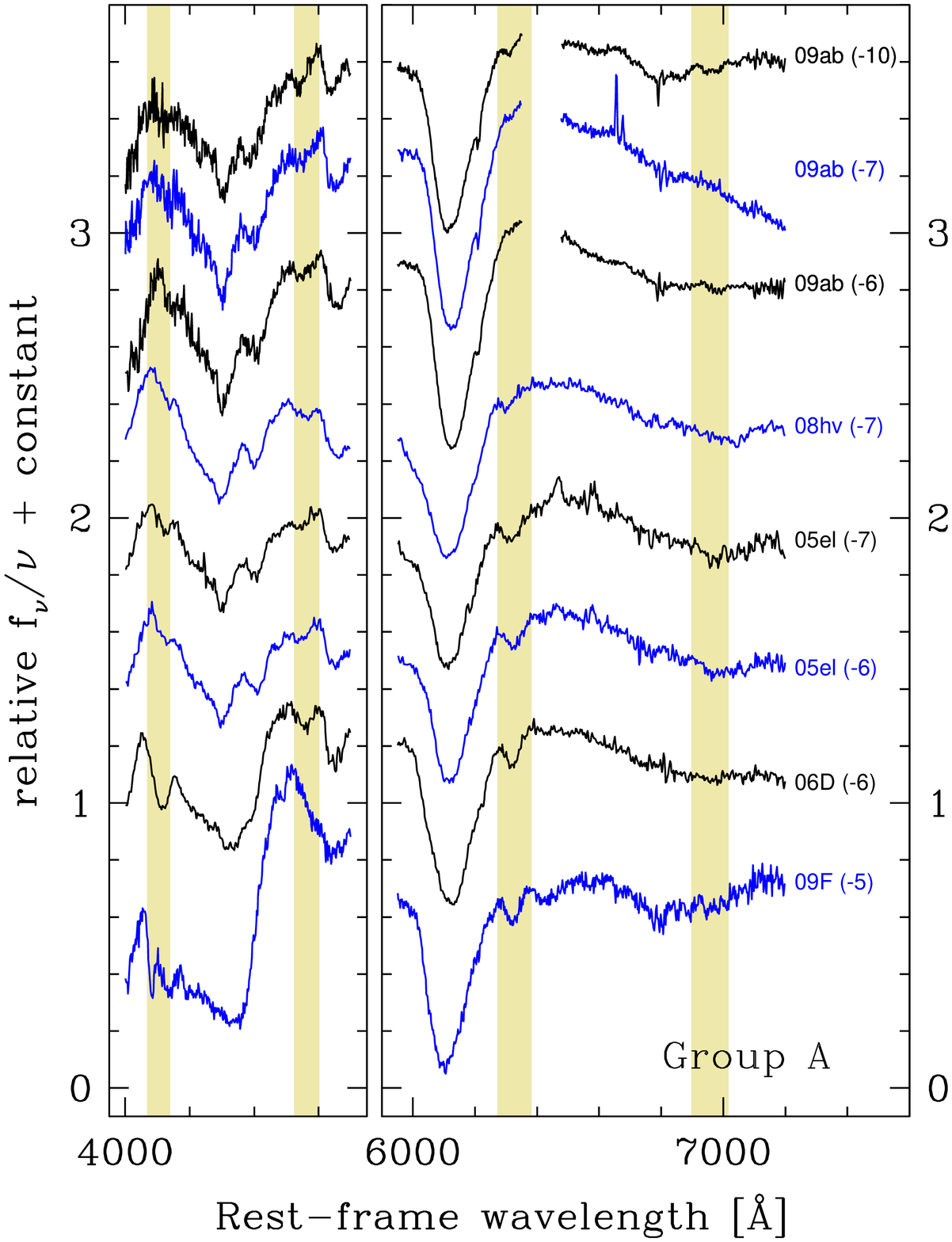}
\caption{Spectra of all the \sneia\ classified in group A (with \ion{C}{2}
  $\lambda$6580 absorption) at the wavelength
  regions of the \ion{C}{2} lines. The
  shaded bands mark the position of four \ion{C}{2} lines blue-shifted
  between 9000 and 14000 km s$^{-1}$. For clarity, the spectra are
  plotted as $f_{\nu}/\nu$. Labels indicate the names of the SNe
  and their age in days relative to $B$-band maximum
  light. The SNe are sorted from top to bottom in increasing
  order of \dm. The gaps 
    in the spectra of SN~2009ab ({\em right panel}) are due to a
    separation between two detectors.\label{fig:CA}} 
\end{figure}

\begin{figure}[htpb]
\epsscale{1.0}
\plottwo{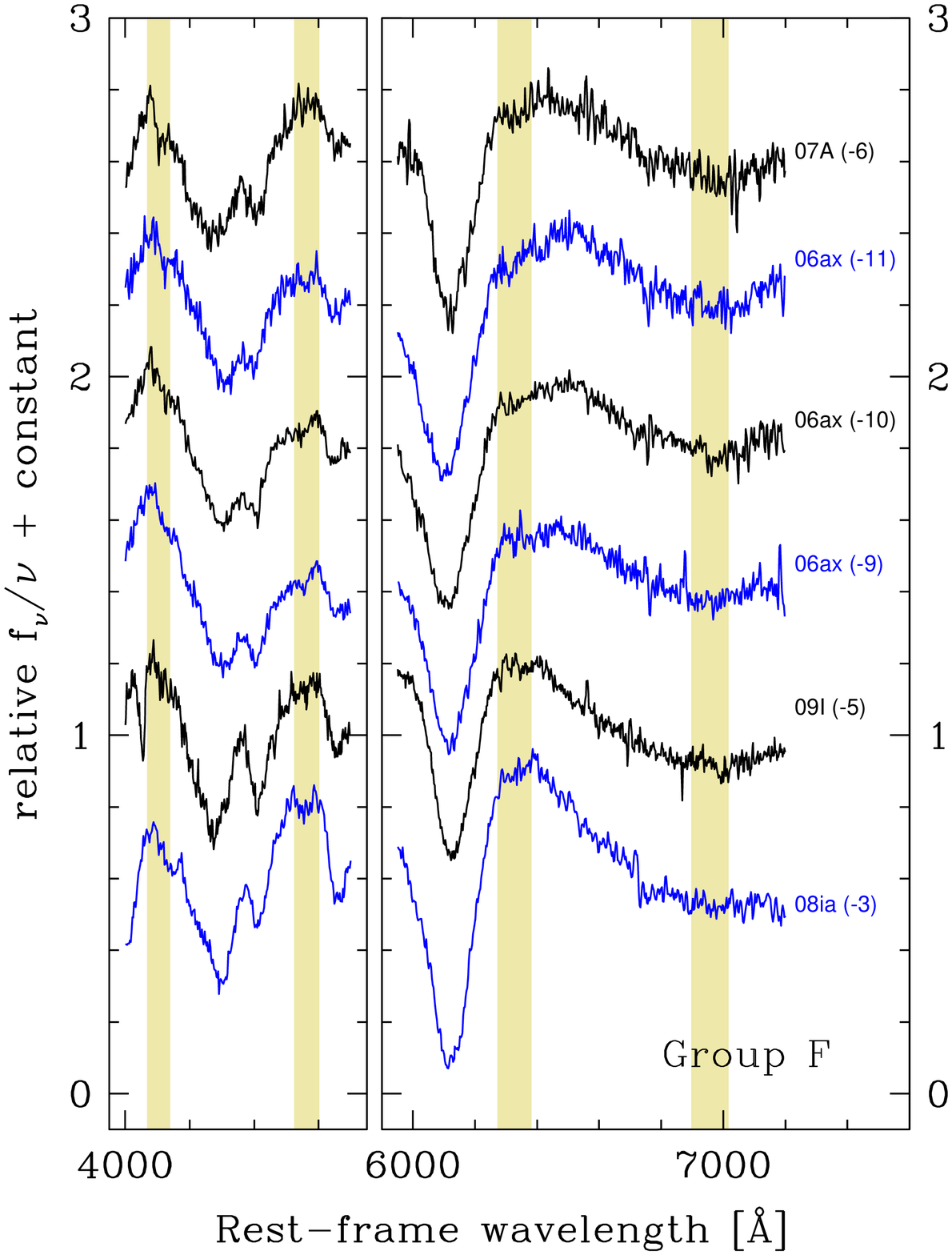}{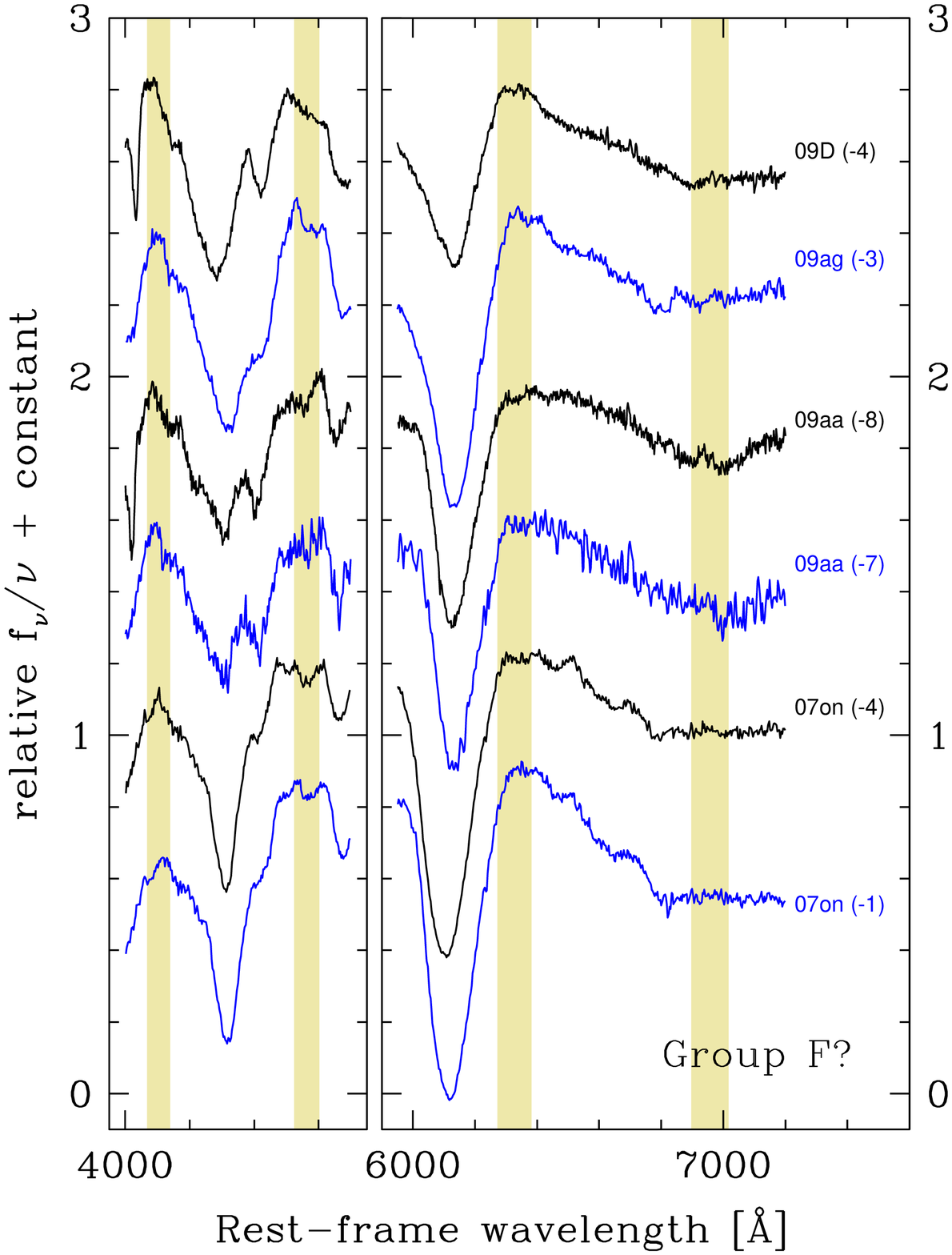}
\caption{({\em Left}) Spectra of all \sneia\ classified in group F (with
  flat \ion{Si}{2} emission) at the wavelength
  regions of the \ion{C}{2} lines. The
  shaded bands mark the position of four \ion{C}{2} lines blue-shifted
  between 9000 and 14000 km s$^{-1}$. For clarity, the spectra are
  plotted as $f_{\nu}/\nu$. Labels indicate the names of the SNe
  and their age in days relative to $B$-band maximum
  light. The SNe are sorted from top to bottom in increasing order of
  \dm. ({\em Right}) 
  The same for \sneia\ with F? classification. \label{fig:CF}} 
\end{figure}

\begin{figure}[htpb]
\epsscale{1.0}
\plottwo{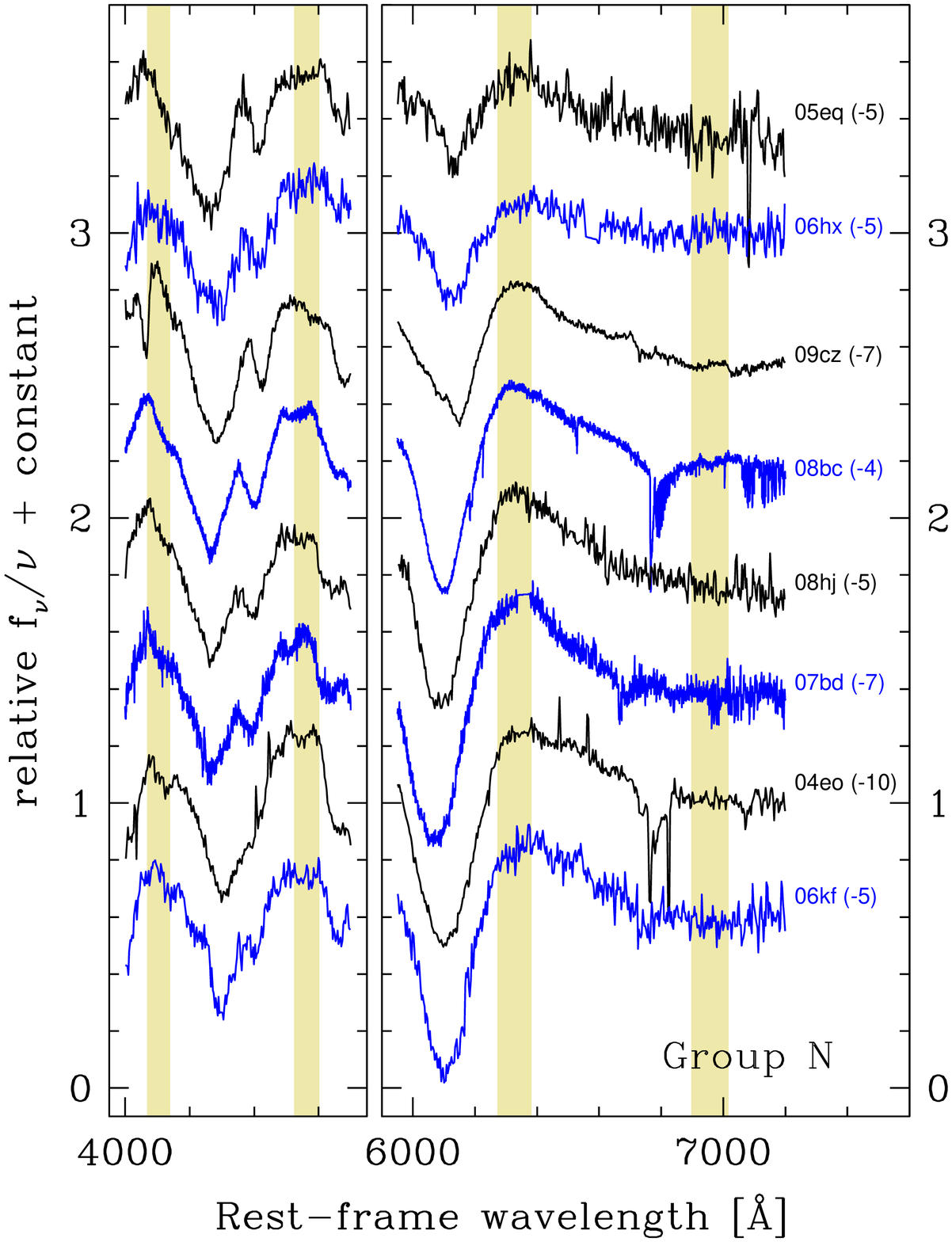}{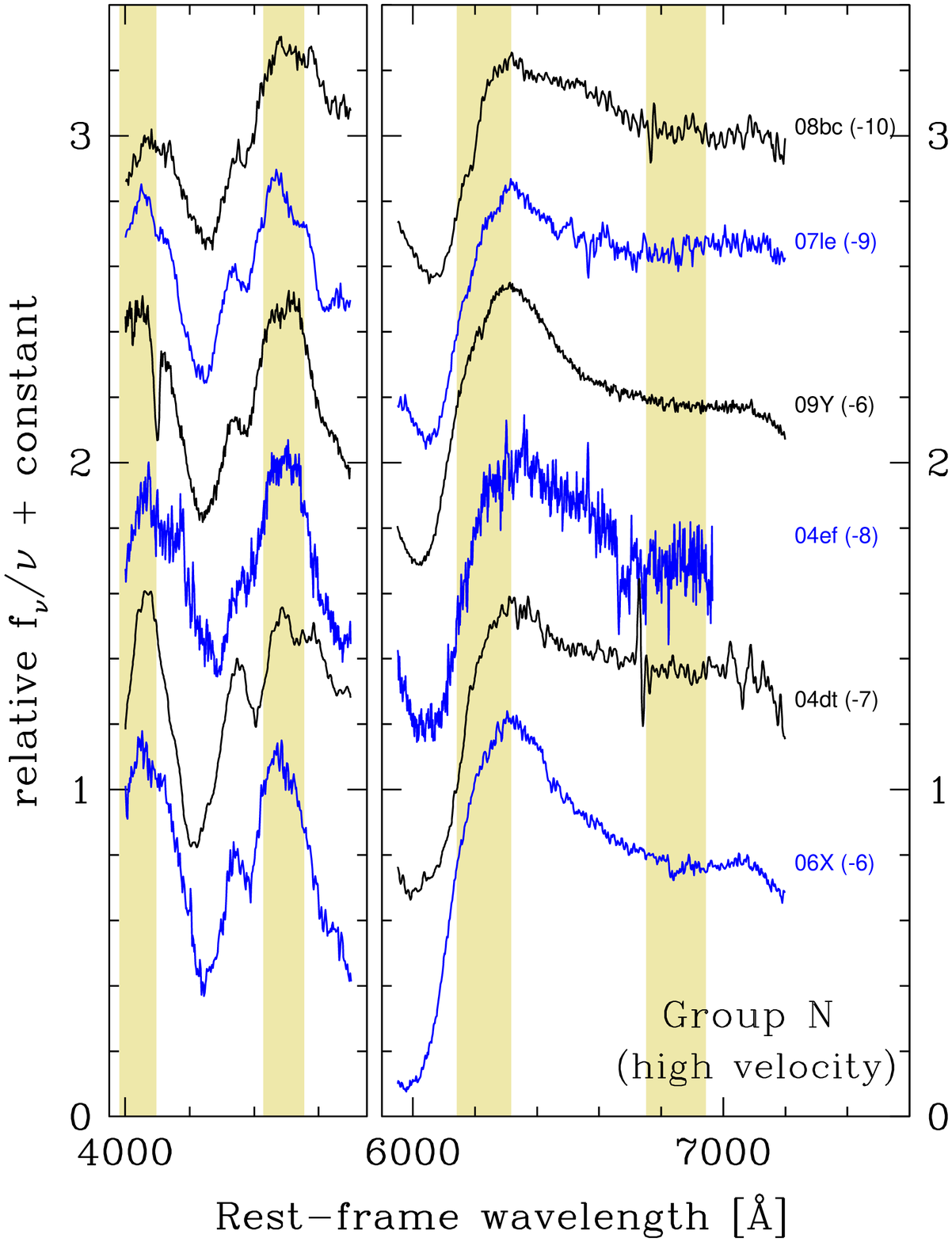}
\caption{({\em Left}) Spectra of all \sneia\ classified in group N
  (without signs of \ion{C}{2}
  $\lambda$6580) at the wavelength
  regions of the \ion{C}{2} lines. The
  shaded bands mark the position of four \ion{C}{2} lines blue-shifted
  between 9000 and 14000 km s$^{-1}$. For clarity, the spectra are
  plotted as $f_{\nu}/\nu$. Labels indicate the names of the SNe
  and their age in days relative to $B$-band maximum
  light. The SNe are sorted in increasing order of \dm. ({\em Right})
  The same for \sneia\ with N classification and \ion{Si}{2} expansion
  velocity greater than 13500 km s$^{-1}$. In this case, the
  shaded bands mark the position of \ion{C}{2} lines blue-shifted
  between 12000 and 20000 km s$^{-1}$. The spectra are sorted from top
  to bottom in 
  increasing order of \ion{Si}{2} expansion velocity. \label{fig:CN}} 
\end{figure}

Visual inspection of the spectra allowed us to divide the
present SN sample into three main groups. First, those SNe which show
no apparent absorption but rather a clear emission component of the \ion{Si}{2}
$\lambda$6355 line. They are marked with ``N'' (i.e., ``no detection'') in
Table~\ref{tab:sne}. Objects that do show a clear absorption feature,
usually centered at $\approx$6320 \AA\ are indicated with an
``A'' (i.e., ``absorption'') in Table~\ref{tab:sne}. An intermediate
case is that of SNe which show no clear absorption but an apparently
suppressed \ion{Si}{2} $\lambda$6355 emission component. Because of
the flat emission profile, these are labeled ``F'' (i.e., ``flat
emission'') in Table~\ref{tab:sne}. We note that the distinction
  between groups A and F is not a quantitative one, and the
  specific classification may depend in each case on the quality of the
  available data. Nevertheless, this does not affect the results
  of the forthcoming analysis. Finally, for those 
cases which are inconclusive ---typically due to data with low
signal-to-noise ratio--- we have added a ``?'' to the identification. In
the cases of SNe~2007S, 2008ar, 2008gp, and 2009ad, the available
spectra do not cover the wavelength range under scrutiny and thus we are not
able to classify them according to the presence of the putative \ion{C}{2} 
$\lambda$6580 line. Figure~\ref{fig:CA} shows the
spectra which belong to group A in the wavelength regions of the four
\ion{C}{2} lines. Figure~\ref{fig:CF} shows the spectra with flat \ion{Si}{2}
$\lambda$6355 emission. In the left panel we include SNe in the F
group while in the right panel objects are classified as ``F?''. As a
comparison, several ``non-detections'', i.e SNe in group N, are shown
in Figure~\ref{fig:CN}. 

We have subsequently studied the spectral regions corresponding to the
other three \ion{C}{2} lines in search of similar features which may
help consolidate the identification with this ion. Although in the
case of some A-type objects other absorptions at the right locations
can be seen (Figure~\ref{fig:CA}), these are usually very weak and it
is difficult to clearly associate them with \ion{C}{2}. In particular,
the two lines 
on the blue part of the spectrum are subject to severe blending into
stronger lines, mainly of \ion{Mg}{2}, \ion{Si}{2}, \ion{S}{2} and
  \ion{Fe}{2}. Moreover, an absorption 
matching the expected position of the \ion{C}{2} $\lambda$4745 line is
seen in most of the spectra, irrespective of the SN belonging to the
A, F or N types. The absorption due to \ion{C}{2}
$\lambda$7234, depending on the redshift and expansion velocity, may
lie near the 6900 \AA\ telluric band, which may affect the detection
if the atmospheric feature is not accurately removed. 

\begin{figure}[htpb]
\epsscale{0.8}
\plotone{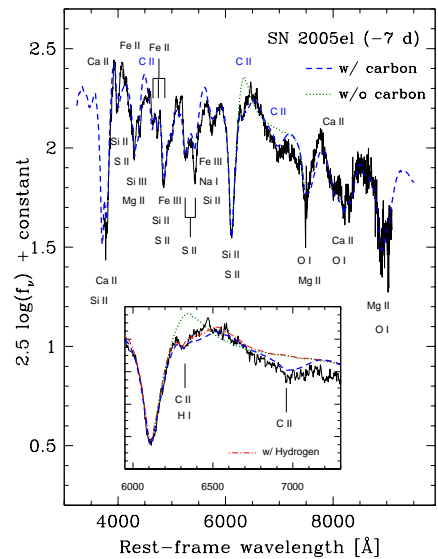}
\caption{Spectrum of SN~2005el at $-7$ days ({\em black solid line})
  which belongs to group A, and a matching SYNOW calculation ({\em
    blue dashed line}). The SYNOW synthetic spectrum without carbon is
  shown for comparison near the \ion{C}{2} $\lambda$6580 and
  $\lambda$7234 lines ({\em green dotted line}). The labels indicate the 
  main ions which contribute to the formation of the observed
  features. The most important contributors for each
  feature are labeled closest to the spectrum. Labels above the
  spectrum indicate that the ion was located ``detached'' from the
  photosphere. ({\em Inset}) Blow-up of the region around \ion{C}{2}
    $\lambda$6580 and $\lambda$7234. The SYNOW spectrum
  containing \ion{H}{1} instead of \ion{C}{2} is also shown ({\em red
    dotted line}). The expected locations of 
  \ion{C}{2} and H$\alpha$ lines are indicated. While \ion{H}{1} can
  reproduce the absorption at $\approx$6300 \AA, \ion{C}{2} produces a
  more solid identification by reproducing both absorptions at
  $\approx$6300 and $\approx$7000 \AA. \label{fig:Csyn05el}} 
\end{figure}

\begin{figure}[htpb]
\epsscale{0.8}
\plotone{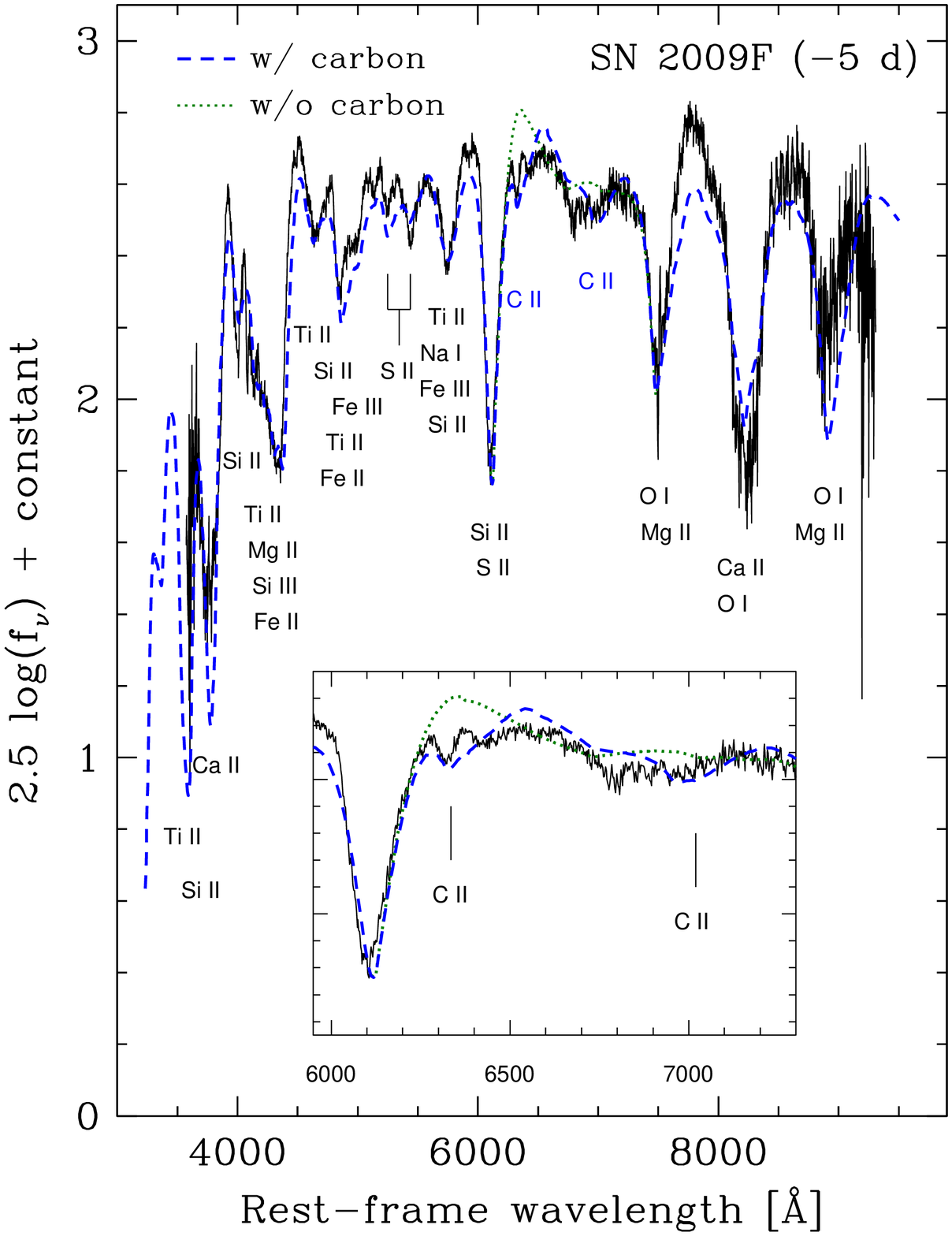}
\caption{Spectrum of the ``Cool'' SN~2009F at $-5$ days ({\em
      black solid line}) 
  which belongs to group A, and a matching SYNOW calculation ({\em
    blue dashed line}). The SYNOW synthetic spectrum without carbon is
  shown for comparison near the \ion{C}{2} $\lambda$6580 and
  $\lambda$7234 lines ({\em green dotted line}). See
  Figure~\ref{fig:Csyn05el} for a description 
  of the labels. ({\em Inset}) Blow-up
  of the region around \ion{C}{2} 
    $\lambda$6580 and $\lambda$7234. The expected locations of 
  \ion{C}{2} lines are indicated. \label{fig:Csyn09F}} 
\end{figure}

\begin{figure}[htpb]
\epsscale{0.8}
\plotone{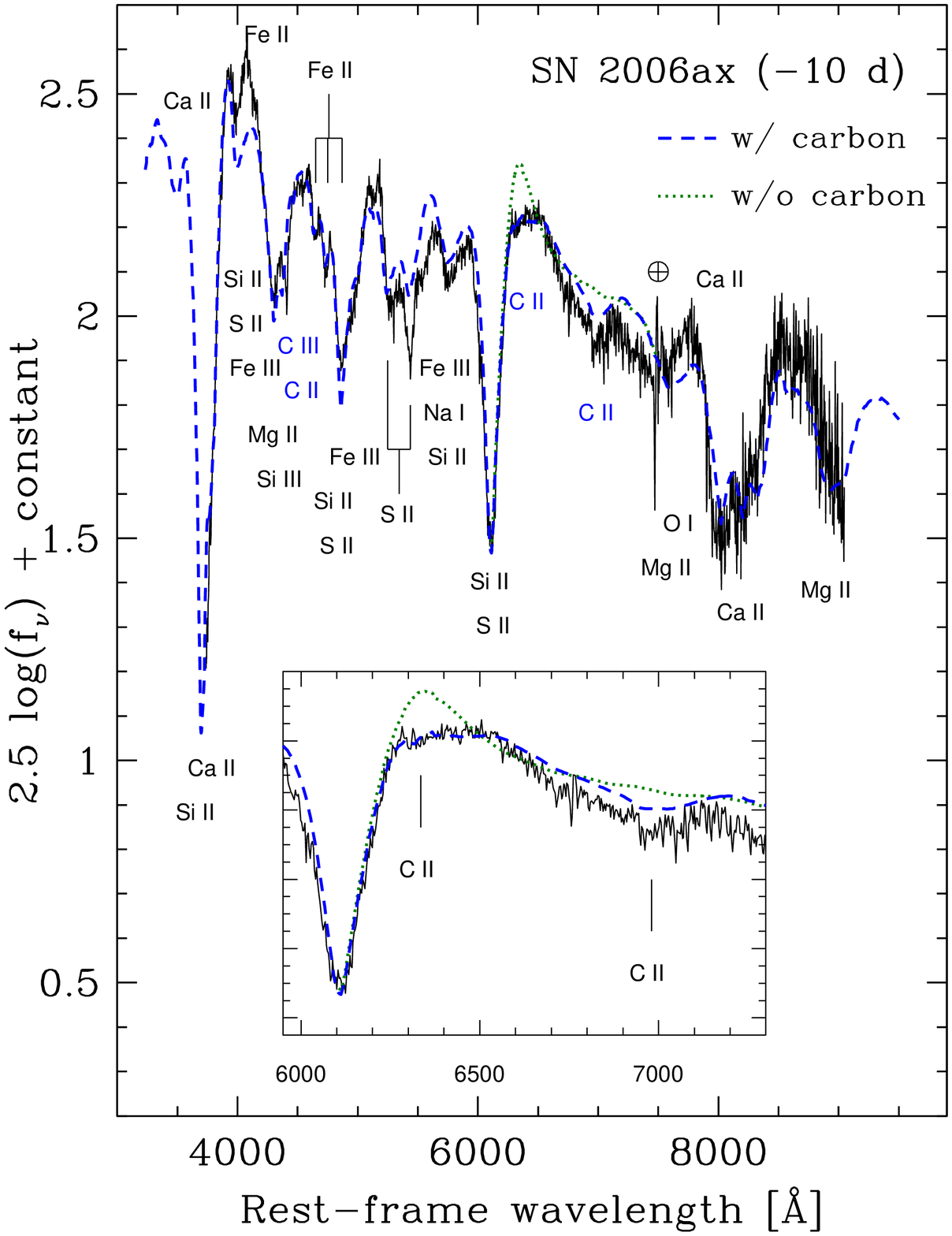}
\caption{Spectrum of SN~2006ax at $-10$ days ({\em black solid line})
  which is classified in group F, and a matching SYNOW calculation
  ({\em blue dashed line}). The SYNOW synthetic spectrum without carbon is
  shown for comparison near the \ion{C}{2} $\lambda$6580 and
  $\lambda$7234 lines ({\em green dotted line}). See
  Figure~\ref{fig:Csyn05el} for a description 
  of the labels. ({\em Inset}) Blow-up
  of the region around \ion{C}{2} 
    $\lambda$6580 and $\lambda$7234. The expected locations of 
  \ion{C}{2} lines are indicated.\label{fig:Csyn06ax}}
\end{figure}

\begin{figure}[htpb]
\epsscale{0.8}
\plotone{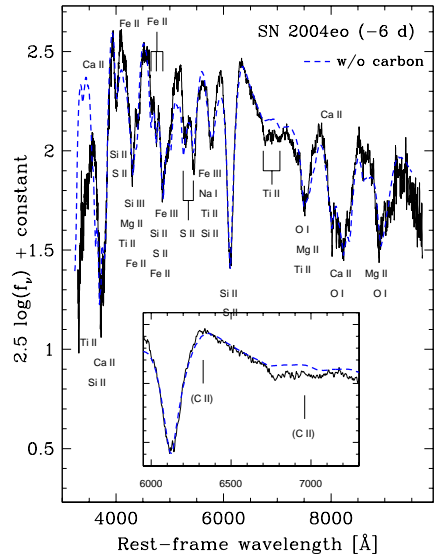}
\caption{Spectrum of SN~2004eo at $-6$ days ({\em black solid line})
  which belongs to the N class, and a matching SYNOW calculation
  without carbon ({\em blue dashed line}). See
  Figure~\ref{fig:Csyn05el} for a description 
  of the labels. ({\em Inset}) Blow-up
  of the region around \ion{C}{2} 
    $\lambda$6580 and $\lambda$7234. The expected locations of 
  \ion{C}{2} lines at about 1000 km s$^{-1}$ above the \ion{Si}{2}
  velocity are indicated. The same as
  above for the spectrum of SN~2004eo at $-6$ days which belongs to
  the N class. \label{fig:Csyn04eo}}
\end{figure}

A more careful identification of spectral lines can be carried out
using the highly parameterized supernova synthetic spectrum
code SYNOW \citep[see][and references therein]{branch02}. 
Examples of synthetic spectra compared with observations are shown in
Figures~\ref{fig:Csyn05el} through \ref{fig:Csyn04eo}. Due to the
approximations involved in SYNOW 
(spherical symmetry, sharp photosphere, LTE, Sobolev line formation,
etc.), the synthetic spectra are not expected to produce a perfect
match to the observed flux and line profiles. Nevertheless, SYNOW is
useful for identifying spectral features with the species which produce
them, specifically by reproducing the location of observed absorptions
with predicted lines of a given ion. 

Table~\ref{tab:synow} provides a summary of the main parameters used
to produce the synthetic spectra shown in
  Figures~\ref{fig:Csyn05el} through \ref{fig:Csyn04eo}. In order to
simplify the 
analysis, several other parameters have been left fixed. In
particular, we have adopted a power law of index 8 for the dependence
of the optical depth on velocity, up to 30000 km s$^{-1}$. We have also
fixed the excitation temperature to $T_{\mathrm{exc}}=$ 12000 K for
all species.

Figures~\ref{fig:Csyn05el} and \ref{fig:Csyn09F} show the cases
  of two objects in group A, namely SNe~2005el and 2009F at $-7$ and
  $-5$ days respectively. These 
SNe show different \dm\ values and correspondingly distinct
  spectroscopic properties. While SN~2005el is a normal event,
  SN~2009F presents fast-declining light curves and belongs to the
  ``Cool'' or ``CL'' type in the scheme of \citet{branch06}. The
  latter SN shows a relatively red spectrum and strong absorptions due
  to \ion{Ti}{2} and \ion{O}{1}, as compared with normal \sneia.  
The synthetic spectra provide satisfactory matches to the
observations, with small deviations which are not relevant to our
  analysis of carbon features. The ions producing the main spectral
features are labeled (see Table~\ref{tab:synow}). In some cases,
``detached'' components of \ion{Ca}{2} and \ion{Fe}{2} have been
  included to reproduce high-velocity (HV) absorption features
  \citep{mazzali05}.

In the case of SN~2005el (see Figure~\ref{fig:Csyn05el}),
\ion{C}{2} can be invoked to reproduce two 
absorptions observed at about 6300 and 7000 \AA. The most evident of
the two is \ion{C}{2} $\lambda$6580 which appears sharper in the data
than in the calculations. The assumed distribution of \ion{C}{2} is
slightly ``detached'' by 500 km s$^{-1}$ above the photosphere in
order to reproduce the location of the lines. The synthetic
spectrum without \ion{C}{2} (dotted line) fails to reproduce those
two features. The identification of two absorptions provides
  extra support to the detection of carbon (see Section~\ref{sec:H?}
  for a comparison with other possible identifications). In the blue,
only \ion{C}{2} $\lambda$4745 produces 
a detectable absorption in the synthetic spectrum. This line roughly
matches an observed absorption, although 
the identification is not clear due to the presence of \ion{S}{2} and
\ion{Fe}{3} lines. Moreover, the synthetic spectrum in that
region is dominated by an emission, which makes the identification
even more doubtful. As will be shown below in the case of SN~2006ax,
this emission could tentatively be suppressed by \ion{C}{3}. No
evident signature of \ion{C}{2} $\lambda$4267 is seen in either the observed
or synthetic spectra at the expected position.  

We refer the reader to a similar analysis done by
\citet{thomas07} on SN~2006D which is also in group A according to our
classification.
The authors discuss whether non-LTE effects may modify the relative
strength of \ion{C}{2} lines to reproduce a strong and sharp
absorption at about 4100 \AA\ which is not identifiable with any other
species.

The case of SN~2009F (see Figure~\ref{fig:Csyn09F}) is similar
to that of SN~2005el. As a 
consequence of SN~2009F being of the ``Cool'' type, the main
differences in the synthetic calculations are a lower black-body
temperature, and 
larger \ion{O}{1} and \ion{Ti}{2} optical depths. Additionally, no
ion ---including \ion{C}{2}--- is detached from the photosphere. Again
the synthetic spectrum with \ion{C}{2} provides a better match to the
observations than the one without carbon, in particular by
  reproducing observed absorptions near 6300 and 7000 \AA, although
  the apparent width of the \ion{C}{2} $\lambda$6580 line is too wide
  in the calculations. The identification of \ion{C}{2} lines in the blue part
of the spectrum is even more difficult than in the case of SN~2005el
due to the presence of strong \ion{Ti}{2} lines.

Figure~\ref{fig:Csyn06ax} shows the case of an F spectrum of SN~2006ax
obtained 10 days prior to maximum light. In this case, even if the
absorption feature is not evident, a comparison between the synthetic
spectra with and without carbon clearly shows that the presence of
\ion{C}{2} is a plausible explanation for the
flat emission component of the \ion{Si}{2} $\lambda$6355 line. A
  weak absorption near 7000 \AA\ can be associated with \ion{C}{2}
  $\lambda$7234, as in the cases of SNe~2005el and 2009F.
We thus assume a common origin for the 
spectral properties of A and F spectra in the range around 6300
\AA. 

Note also in Figure~\ref{fig:Csyn06ax} that we have tentatively included
\ion{C}{3} in the synthetic spectrum. This adds an absorption at about
4500 \AA\ due to \ion{C}{3} $\lambda$4650 which serves to suppress the
emission components of \ion{Mg}{2} and \ion{Si}{3} lines to the
blue. As mentioned above, \ion{C}{3} can also be invoked to improve
the fit in this wavelength region in the case of SN~2005el, as shown
in Figure~\ref{fig:Csyn05el}. Although the presence of \ion{C}{3}
  $\lambda$4650 has been considered previously in the literature
  \citep[see][]{garavini04,branch07,parrent11}, its identification is
  only tentative because it would require a very high
  temperature to excite this transition ($29.5$ eV). It should be
noted that more elaborate models, for example those applied to the
normal SN~2003du by \citet{tanaka11}, do not involve the presence of
\ion{C}{3} lines even when some carbon is included in the ejecta.

For comparison, in Figure~\ref{fig:Csyn04eo} we show
the case of an N type spectrum. In this example, we show 
the spectrum of SN~2004eo obtained at $-6$ days along with a matching
SYNOW spectrum. We see that, as opposed to the A and F cases, the
match around 6300 \AA\ is good without the need to invoke
\ion{C}{2}. In a case like this we can confidently say that carbon is
not detected. However, as shown in Section~\ref{sec:Cdet}, in
other cases classified as N or F?, carbon features may be hidden by
noise or blending. 

\subsection{Detection Statistics}
\label{sec:Cdet}

Assuming the identification of the 6300 \AA\ feature in A and F SNe is
\ion{C}{2}, we see a remarkably large incidence of unburned
material in the current sample. If we consider the objects with
spectroscopy obtained before or at $-4$ days ---the age of the oldest
A spectrum---, from a total of 32 normal \sneia\
we have nine objects in group A and three in group F. This means 38\% of
the SNe are in the A or F groups, and 28\% in A. Apart from the recent
work by \citet{parrent11} who find a similar detection fraction of 30\%,
previous claims for carbon in \sneia\ spectra have been sporadic
\citep[see][]{thomas07,branch07}. As shown below, we think this
has to do with the relative scarcity of early-time spectra.

Moreover, we note that the fractions derived above are lower limits
because we count as non-detections several cases where the data are
not suitable to fully discard the presence of carbon. Below we 
present three different effects which can work against the detection of
\ion{C}{2} lines: noise, expansion velocity, and age. 

\subsubsection{Noise}
\label{sec:noise}

\begin{figure}[htpb]
\epsscale{1.0}
\plotone{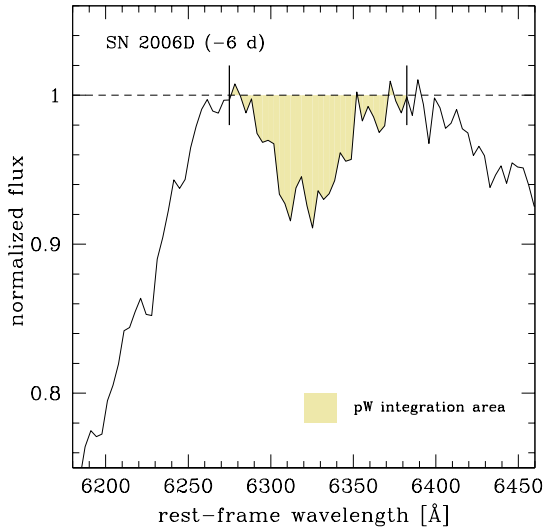}
\caption{Example of a pseudo-equivalent width measurement (\ew) for
  the \ion{C}{2} $\lambda$6580 line. A pseudo-continuum is defined as
  a straight line traced between the flux peaks
  on each side of the absorption feature ({\em vertical lines}). The
  spectrum is then normalized by this pseudo-continuum, and the
  \ew\ is computed as the area of the normalized absorption ({\em 
  shaded region}). \label{fig:ewex}}    
\end{figure}

\begin{figure}[htpb]
\epsscale{1.0}
\plotone{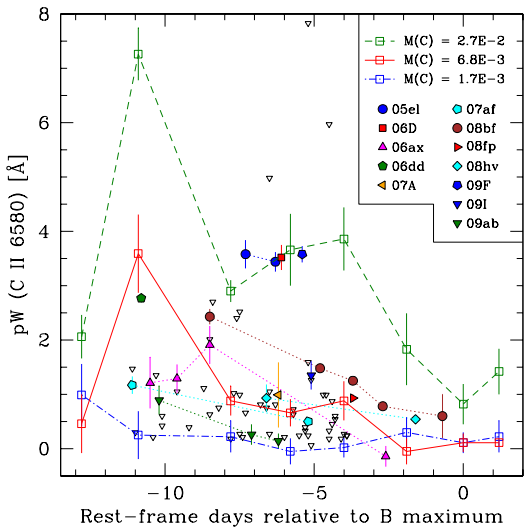}
\caption{Pseudo-equivalent width of the absorption at $\approx$6300
  \AA\ which is attributed to \ion{C}{2} $\lambda$6580. Filled symbols
  are used for \sneia\ in A and F groups. Open triangles mark 3-$\sigma$
  upper limits for the rest of the spectra in the current
  sample. Open squares linked with lines show the results for
  synthetic spectra computed for different masses of carbon in the
  velocity range 10500 $< v < $ 15000 km s$^{-1}$ (see
  Section~\ref{sec:MC}).\label{fig:ew}}   
\end{figure}

Clearly, noise in the data can hide weak features such as the ones
under study here. In order to quantify the detectability of the
absorption at 6300 \AA, we have performed equivalent width
measurements. Strictly speaking, since the location of the actual
continuum is unknown, these are {\em pseudo}-equivalent widths
(\ew). We simply trace a straight line along the absorption
feature to mimic the continuum flux. An example of a
  \ew\ measurement is shown in Figure~\ref{fig:ewex}. The \ew\ is
  computed as the area of the absorption feature after normalizing the
  spectrum by a local pseudo-continuum.
In cases when no absorption is
seen, we derive an uncertainty in the \ew\ value in the
region around 6300 \AA\ in order to set a 3-$\sigma$ upper limit. The
measured values and upper limits are listed in Table~\ref{tab:spec},
and shown in Figure~\ref{fig:ew}. 

We can see that, in general, \ew\ $\gtrsim 1$ \AA\ for objects in the A and F
groups, and that most of the 3-$\sigma$ upper limits of SNe with no \ion{C}{2}
detections lie below that value. However, in several cases the upper
limits are comparable to or even larger than the measured \ew\ of
objects in groups A and F. This implies that some of the \sneia\ with
carbon may have gone undetected because of noise in the spectra.

\subsubsection{Expansion Velocity}
\label{sec:expvel}

The higher the expansion velocity, the wider the spread of spectral
features in wavelength space, and thus the larger the impact of line
blending. In particular, the hypothetical 6300 \AA\ absorption, which
is much weaker than its neighboring \ion{Si}{2} $\lambda$6355 line, would
become increasingly blended into the red absorption wing of the
latter. This would make the detection of carbon problematic when
expansion velocities are high.

\begin{figure}[htpb]
\epsscale{0.8}
\plotone{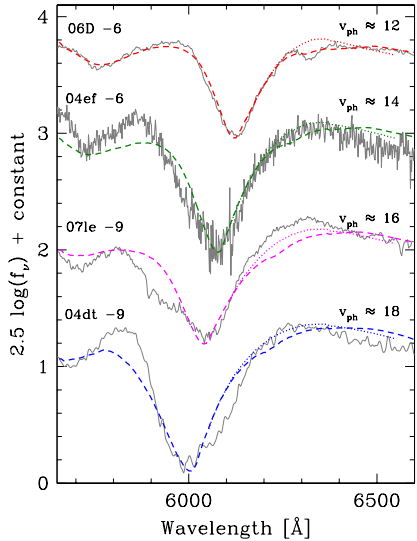}
\caption{The effect of Doppler blending on the possible detection of
  \ion{C}{2} $\lambda$6580. Four observed spectra are plotted ({\em
    gray continuous lines}) with
  the labels on the left side indicating the SN name and epoch in
  days relative to $B$ maximum. SYNOW calculations accompanying each
  spectrum show the cases with ({\em dashed lines}) and without
  ({\em dotted lines}) \ion{C}{2}. The top synthetic spectrum was
  produced to fit the 
  complete observed spectrum of SN~2006D at $-6$ days. This fiducial
  spectrum was used to produce the other synthetic spectra with
  different photospheric velocities. The labels on the right indicate
  this velocity in units of $10^3$ km s$^{-1}$. \label{fig:synvel}}
\end{figure}

To illustrate this problem, plotted in Figure~\ref{fig:synvel} are
synthetic spectra obtained with SYNOW using 
different photospheric velocities. For comparison, data of different
SNe with approximately matching expansion velocities are shown along with the
synthetic spectra. The top synthetic spectrum was tailored to match the
observed one of SN~2006D at $-6$ days. We note that the \ion{C}{2}
$\lambda$6580 absorption in the synthetic spectrum is wider than the
observed one, but it reproduces well the observed
  intensity. This SYNOW spectrum is taken as the fiducial 
model to produce the rest of the synthetic spectra shown in the
figure. In each case we 
have only varied the photospheric velocity and optical depth of
\ion{Si}{2} in order to match the location and strength of the
\ion{Si}{2} $\lambda$6355 absorption. Note that we have not attempted
to produce an accurate match of each spectrum, but we have only tried
to illustrate the effect of Doppler blending. 

It is clear from Figure~\ref{fig:synvel} that, when the expansion
velocity exceeds roughly 15000 km s$^{-1}$, a \ion{C}{2} absorption as strong
as or weaker than the one of SN~2006D would
be very difficult to detect. In the case of SN~2004ef at $-6$ days, the
signal-to-noise ratio of the data makes the detection of the
hypothesized \ion{C}{2} line barely possible. For SN~2007le at $-9$
days, we can observe some ``wiggles'' on the \ion{Si}{2} $\lambda$6355
emission profile which may be compatible with a \ion{C}{2}
absorption. These weak features, along with the complex 
structure of the \ion{Si}{2} absorption, disappear by the following
observation of SN~2007le obtained at $-4$ days. This led to our
classification of this object as ``F?''.

The right-hand panel of Figure~\ref{fig:CN} shows several cases of
non-detections of \ion{C}{2} for \sneia\ with \ion{Si}{2} velocities above
13500 km s$^{-1}$. In some of these cases, specifically for SNe~2004dt
and 2006X, the expansion velocities are so large that they would hide
the weak \ion{C}{2} absorption observed in the case of type A \sneia. We
note that the nature of SN-Ia explosions conspires with this effect
against the detection of carbon. This is because expansion velocities
are the largest at the earliest phases, which implies that the line
blending effect is worst at times when the plausible \ion{C}{2}
absorption is expected to be strongest (see Section~\ref{sec:age}).

As shown in Section~\ref{sec:spprop}, none of the spectra
  classified as A or F show a \ion{Si}{2} $\lambda$6355 velocity
  greater than $\approx$12500 km s$^{-1}$. According to the
  brief analysis based on Figure~\ref{fig:synvel}, the detection of
  the \ion{C}{2} $\lambda$6580 line could be possible
   for velocities above that value and up to $\approx$15000
  km s$^{-1}$. This may indicate that the lack of carbon detections
  among spectra within this range of expansion velocities is not
  exclusively due to the effect of Doppler blending.

\subsubsection{Age}
\label{sec:age}

All the claimed detections of \ion{C}{2} in the spectra of normal
\sneia, as the ones presented in this paper and in previous
publications \citep{thomas07,branch07,parrent11}, 
correspond to early-time observations, most commonly between one and
two weeks before maximum light. This is mainly because carbon is
  expected to be present in the outer region of the ejecta which are
  visible at early phases. This is consistent with the temporal
  decrease in the strength of the absorption at $\approx$6300 \AA, as can be
seen in Figure~\ref{fig:ew}. The case of SN~2008bf,
shown on the left panel of Figure~\ref{fig:CA}, is a clear example of
this evolution between $-9$ and $-3$ days. An additional spectrum of
this SN, obtained at $-1$ day, is classified as N (see Table~\ref{tab:spec}). 

\begin{figure}[htpb]
\epsscale{1.0}
\plotone{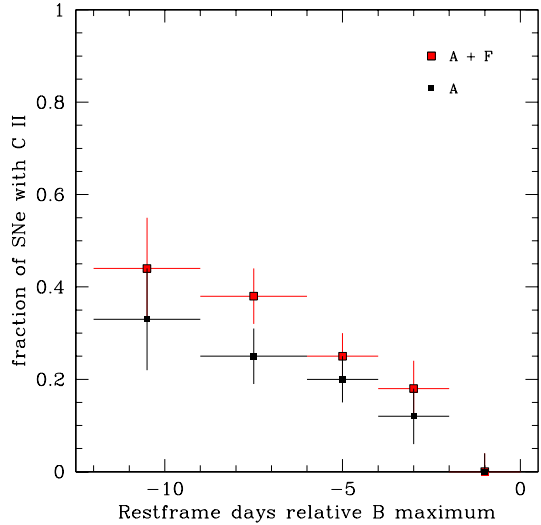}
\caption{Fraction of SNe with \ion{C}{2} as a function of
  age. Values show the ratio of SNe with spectra in the A ({\em
    solid black}) or A+F ({\em shaded red}) groups to the total number
  of SNe with spectra covering the region around 6300 \AA. Horizontal
  error bars show the size of the epoch boxes used for 
  computing statistics. Vertical error bars indicate the expected
  variation of the fraction for an assumed error of $\pm$1 SN with
  carbon. \label{fig:Cfrac}} 
\end{figure}

Sample statistics further support this picture. Figure~\ref{fig:Cfrac}
shows the fraction of carbon detections as a function of age for the
current sample. The fractions are simply the ratio of the number of
SNe in group A (or A and F) to the total amount of objects whose
spectra cover the region around 6300 \AA. As we go from $-11$ to $-1$
day, the fraction decreases monotonically from about 30\% (40\% for
A+F) to zero detections. 

In the next section we show that some models predict a weaker
absorption at $\approx$6300 \AA\ before $-11$ days, which may lead to
a lower detection fraction prior to that epoch. This highlights
the importance of early observations in determining the
  incidence of unburned material in the ejecta of \sneia.

\subsection{The Mass of Carbon}
\label{sec:MC}

In an attempt to put constraints on the amount of unburned material
which is necessary to explain the observed spectral features, we have
computed synthetic spectra using the Monte Carlo code developed by
\citet{mazzali93} and updated by \citet{lucy99} and \citet{mazzali00}. A recent
description of the code and its 
approximations was given in an analysis of SN~2003du by
\citet{tanaka11}. For that standard SN \citep{stanishev07} the authors
find that a mass  
of carbon of $M(C)=6.8 \times 10^{-3} \, M_\odot$ (mass fraction
$X(C)=0.002$) in the velocity range 10500 $< v <$ 15000 km 
s$^{-1}$ is suitable to explain the suppressed emission peak of the
\ion{Si}{2} $\lambda$6355 line observed at $-11$ days. More
carbon is expected to be present in outer layers, but it would not
produce any noticeable effect in the spectrum \citep[see Figure~C1
  of][]{tanaka11}. The authors set an upper limit of $0.016 \,
M_\odot$ for the total mass of carbon in the ejecta of 
SN~2003du. Based on this configuration we produce two
additional models by only changing the mass of carbon in the
velocity range given above to four times below and above the
fiducial value of $M(C)=6.8 \times 10^{-3} \, M_\odot$.
The three models are employed to compute synthetic spectra
at several phases between $-13$ and $+1$ days from maximum light and
thus evaluate the effect on the \ion{C}{2} $\lambda$6580 line. 

Figure~\ref{fig:ew} shows the \ew\ of \ion{C}{2} $\lambda$6580 as a function of
SN phase for the synthetic spectra of different carbon mass. As
expected, the \ew\ generally grows with mass. The range of 
observed \ew\ is well represented by the assumed range of $M(C)$. The
lowest adopted mass ($1.7 \times 10^{-3} \, M_\odot$) would correspond
to SNe with no carbon detection, while the largest mass ($2.7 \times
10^{-2} \, M_\odot$) roughly matches the strongest observed \ion{C}{2}
6580 lines for SNe in group A at about one week before maximum light.

In general, the models confirm the observed fact that the \ew\ of
carbon tends to decrease with phase. The exception to this is the
increase in \ew\ between the modeled spectra at $-13$ and $-11$
days. Close inspection of the synthetic spectra at $-13$ days shows 
a strong emission component of the \ion{Si}{2} line 
which washes away the weak absorption due to \ion{C}{2} and produces a
low \ew\ as compared with that at $-11$ days. To a lower
extent, the \ion{C}{2} \ew\ is additionally reduced by the fact that
at $-13$ days part of the carbon lies below the
photosphere. Furthermore, a relatively larger expansion velocity at
day $-13$ enhances line blending, thus contributing to the decrease in
the measured \ew. If this effect were generally valid, one would expect
that the fraction of SNe with carbon detection should {\em decrease}
at phases earlier than roughly $-12$ days. It should be noted,
however, that the adopted model produces a worse match to the observed
spectrum of SN~2003du at $-13$ days than at later phases \citep[see
  Figure~2 of][]{tanaka11}, especially in the wavelength region under
scrutiny here. Unfortunately our data do not cover such early phases
and thus we cannot corroborate the results of the model. More
early-time spectroscopy of \sneia\ is needed to settle this question.

\subsection{Hydrogen?}
\label{sec:H?}

With the aid of SYNOW we have studied alternative identifications for
the absorption around 6300 \AA\ in A-type \sneia. Other ions which may
produce such absorption at expansion velocities near the photospheric
value are \ion{Ne}{1} and \ion{H}{1}. The strongest \ion{Ne}{1} lines
(at 6402 and 6506 \AA) would only reproduce the absorption at
$\approx$6300 \AA\ if the expansion velocity were between 4000 and
9000 km s$^{-1}$. Such velocities are lower than those observed for
\sneia\ before maximum light.

On the other hand \ion{H}{1} is a plausible alternative
\citep{lentz02}. H-rich material may be present by mass transfer from
a companion star in a single-degenerate progenitor system. If the
observed absorption were due to \ion{H}{1} this would rule out the
double-degenerate scenario for a fraction of \sneia.

As can be seen in the inset plot of Figure~\ref{fig:Csyn05el} (dotted lines),
H$\alpha$ {\em at photospheric velocity} is able to reproduce the 6300
\AA\ absorption as well as \ion{C}{2} does. The weaker absorption at
about 7000 \AA\ is, however, not reproduced. This makes the
  identification of \ion{H}{1} less plausible than that of \ion{C}{2}. With
the given optical depth for H$\alpha$, the expected strength of
H$\beta$ is barely detectable, and it is not observed in the data. 

We have computed Monte Carlo synthetic spectra including H in order to
model the observations. For this purpose we utilized the abundance
distribution derived for SN~2003du at $-11$ days as given by
\citet{tanaka11}. We added H with mass fraction up to $0.1$ in the
range of 10500 $<$ $v$ $<$ 15000 km s$^{-1}$ ---i.e., right above
the photosphere. We found that the absorption at 6300 \AA\ for SNe in
the A group can be reproduced with mass fractions $X(H) \sim 0.05$
which in this case translates into a total hydrogen mass of $M(H) \sim
10^{-2} \, M_\odot$. This is a very large amount of hydrogen which is
incompatible with the theoretical picture of \sneia. In the
single-degenerate scenario, models of progenitors with mass 
transfer predict an upper limit of $\sim$$10^{-7} \, M_\odot$, which is
set by the ignition mass of H \citep{nomoto07}. 
In the double-degenerate scenario, an even smaller amount of hydrogen
is expected.

Alternatively, a mass of the order of $0.01 \, M_\odot$ of hydrogen
could be stripped from the companion star in the single-degenerate 
scenario \citep{marietta00,pakmor08}. However, according to these
simulations, the stripped hydrogen is distributed in the inner layers
\citep[v $<$ 4000 km s$^{-1}$;][]{pakmor08}, which are not visible
in early-time spectra.

Based on these observational facts and theoretical predictions, even
if the 6300 \AA\ absorption can be reproduced by hydrogen, we consider this
possibility to be unlikely in favor of the \ion{C}{2} identification. 

\section{SPECTROSCOPIC PROPERTIES}
\label{sec:spprop}

\begin{figure}[htpb]
\epsscale{1.0}
\plotone{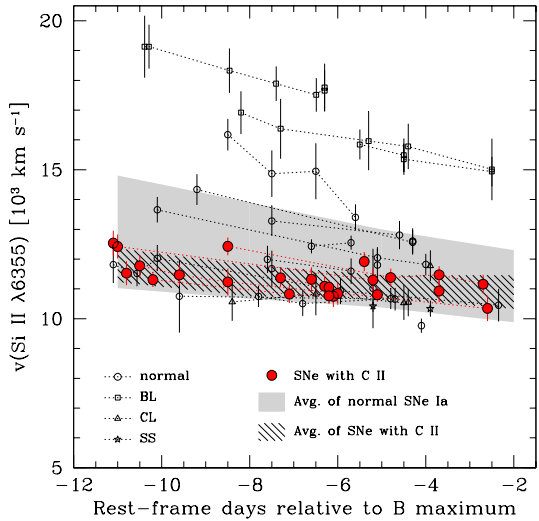}
\caption{Expansion velocities of \ion{Si}{2} $\lambda$6355 as a
  function of epoch relative to $B$ maximum. Open black symbols mark SNe
  without carbon, with different shape for different spectral subtypes
  as given by \citet{folatelli11}. Filled symbols are used for SNe
  in the A and F groups. These symbols are the same as in
  Figure~\ref{fig:ew}. Data points for the same SN are linked with
  dotted lines. The shaded gray region marks the region within
  1$\sigma$ about the average velocity evolution of normal \sneia\ as
  given by \citet{folatelli11}. The yellow hashed region indicates
  the average and 1-$\sigma$ region for SNe with carbon. \label{fig:Sivel}}
\end{figure}

We now compare the spectral properties of the
samples of SNe with and without evidence of carbon. We first examine
the expansion velocities. Figure~\ref{fig:Sivel} shows \ion{Si}{2}
$\lambda$6355 expansion velocities as derived from the blue-shift of the
absorption minimum. We see that the sample of normal \sneia\ with carbon
appear to lie preferentially near the lower edge of the
velocity distribution of the complete sample. \ion{Si}{2} velocities
are below $\approx$12500 km s$^{-1}$ in all of these cases. In terms of the
classification by \citet{benetti05}, none of the SNe with carbon are
of high velocity-gradient (HVG) type. Accordingly, none of the SNe with
carbon belong to the ``Broad Line'' (BL) type as defined by
\citet{branch06} \citep[see also][]{folatelli11}. 
As explained in Section~\ref{sec:expvel}, the lack of carbon detection
when velocities are high may be caused by an observational bias as a
consequence of Doppler blending. However, the fact that in our
simulations \ion{C}{2} lines are expected to be noticeable up to
velocities of $\approx$15000 km s$^{-1}$, suggests that the preference
for low velocities is real and not completely due to the observational
bias. The possible implications of the velocity of carbon on
\sneia\ models are briefly discussed in Section~\ref{sec:dis}.

Signatures of carbon are not exclusively found in normal \sneia. In our
sample, one fast-declining object, SN~2009F, belongs to group A (see
Figure~\ref{fig:Csyn09F}). In the classification scheme 
of \citet{branch06}, this SN is of the ``Cool'' (CL) type. Three other
CL SNe were observed at the range of epochs considered here, with no
evidence of \ion{C}{2} lines. The two SNe of the ``Shallow Silicon''
(SS) type with early-time spectroscopy do not show evidence of carbon.

\begin{figure}[htpb]
\epsscale{1.0}
\plotone{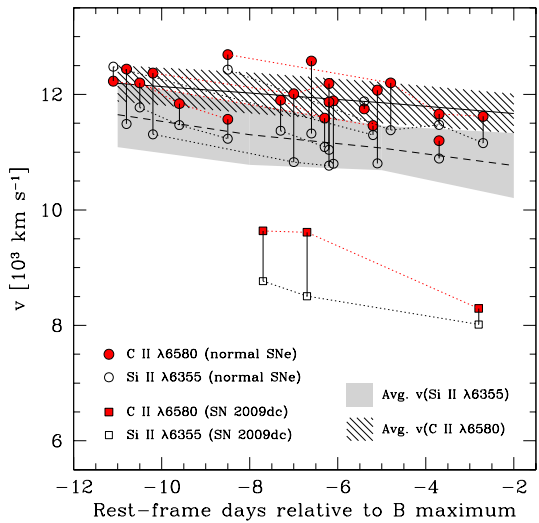}
\caption{Expansion velocities of \ion{C}{2} $\lambda$6580 ({\em filled
  symbols}) and \ion{Si}{2} $\lambda$6355 ({\em open symbols}) as a
  function of epoch relative to $B$ maximum. The dashed line marks
  the average \ion{Si}{2} $\lambda$6355 velocity and the gray shaded
  area indicates the 1-$\sigma$ dispersion. The solid line and hashed
  region mark the average and 1-$\sigma$ dispersion of \ion{C}{2}
  $\lambda$6580 velocities. SN~2009dc is excluded from both
  averages. \label{fig:Cvel}} 
\end{figure}

In Figure~\ref{fig:Cvel} we present the expansion velocities of the
bulk of carbon, measured from the absorption minimum located at
$\approx$6300 \AA, under the assumption that it is due to \ion{C}{2}
$\lambda$6580. On average, the evolution of \ion{C}{2} $\lambda$6580
velocities is parallel to that of \ion{Si}{2} $\lambda$6355
velocities, and lying at $\approx$1000 km s$^{-1}$ above. 
These velocities indicate that carbon appears slightly detached above
the photosphere (see Section~\ref{sec:CIIse}). If carbon were left
unburned by a spherically symmetric explosion, then it would be
expected to only be present in the outermost regions of the
ejecta \citep[e.g.,][predict a minimum velocity of 15000 km
  s$^{-1}$]{nomoto84,iwamoto99}. For 
current one-dimensional deflagration and delayed-detonation models, no
carbon should be left at such low velocities. As discussed in
Section~\ref{sec:dis}, multi-dimensional effects may account for the
presence of carbon deep in the ejecta.

\begin{figure}[htpb]
\epsscale{0.8}
\plotone{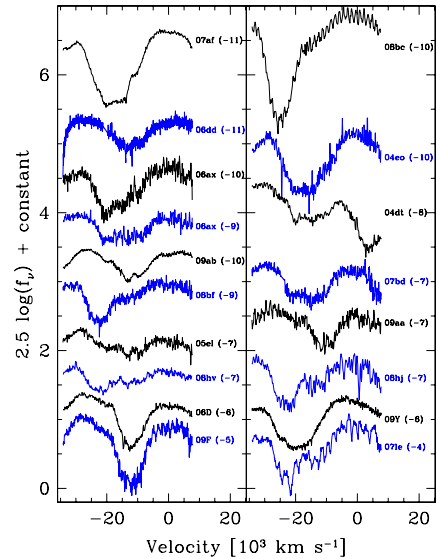}
\caption{Spectra in the region of the \ion{Ca}{2} IR triplet as a
  function of Doppler velocity assuming 8579 \AA\ as the effective
  rest wavelength. In the 
  left panel, spectra classified in A and F groups. In the right
  panel, spectra of type N. The spectra are sorted according to their
  age (which is given in parentheses). We see that cases with and
  without strong high-velocity components ---near 20000 km
  s$^{-1}$--- are present in both groups of
  \sneia. \label{fig:Cahv}}  
\end{figure}

If the detection of carbon is related to prevalent asymmetries in the ejecta,
we would expect to find evidence of other effects of asphericities. In
addition to the distinction between LVG and HVG SNe which was
suggested as evidence of asymmetries \citep{maeda10c},
another possible effect would be the presence of high-velocity
components in the spectra \citep{tanaka06}. As was shown in
Section~\ref{sec:CIIse}, at the SN phases 
studied here, high-velocity components of \ion{Ca}{2} and \ion{Fe}{2}
are commonly seen \citep[see also][]{mazzali05}. Figure~\ref{fig:Cahv}
shows, however, that the presence or absence of high-velocity
\ion{Ca}{2} infrared (IR) triplet absorption is not obviously related
to the presence of carbon. 

\subsection{Super-Chandrasekhar SNe}
\label{sec:sch}

As a comparison we examine the CSP spectra of the  
peculiar SN~2009dc.
This SN exhibits strong carbon features in its pre-maximum spectra,
 which are also present until approximately a week after maximum.
From its spectral properties, slow light curves and exceedingly high
luminosity, it has been suggested SN~2009dc resulted from
a super-Chandrasekhar mass progenitor
\citep{yamanaka09b,tanaka10,silverman11,taubenberger11}. 

In Figure~\ref{fig:Sivel} we compare the expansion velocities of
\ion{Si}{2} $\lambda$6355 and \ion{C}{2} $\lambda$6580 derived from
the spectra of SN~2009dc to our 
sample of normal \sneia. This figure illustrates SN~2009dc has lower
\ion{Si}{2} velocities than the normal objects by $\approx$2000 km
s$^{-1}$. Interestingly, the relation between \ion{C}{2} and
\ion{Si}{2} velocities that was observed for normal \sneia\ also holds for
SN~2009dc. In this case, the bulk of carbon also 
appears to be located above the silicon layer by $\approx$1000--2000 km
s$^{-1}$. From a study of a different sample of \sneia, including a few
super-Chandrasekhar candidates, \citet[][see their Figure~8]{parrent11}
found a similar agreement of the relative location of carbon and silicon
layers. This is an interesting finding because it may indicate that,
even if the progenitors may be very different, the explosion produces
similar relative distribution for the bulk of these elements.

\section{PHOTOMETRIC PROPERTIES}
\label{sec:phprop}

This section presents a comparative analysis of the photometric
properties of \sneia\ with and without signatures of carbon in the
early-time spectra. For this purpose, a sample of 32 normal
  \sneia\ was selected with spectra obtained earlier than four days
before maximum light. Twelve of these objects
belong to groups A and F, which are the ones considered to show carbon
in their ejecta. We base this study on the photometric data published
by \citet{contreras10} and \citet{stritzinger11}. The data were
analysed with the SNooPy package \citep{burns11} in order to derive
light-curve parameters such as times and magnitudes at maximum light,
and $B$-band decline rates \citep[\dm;][]{phillips93}.

\subsection{Decline rate and color}
\label{sec:dmbv}

\begin{figure}[htpb]
\epsscale{1.0}
\plottwo{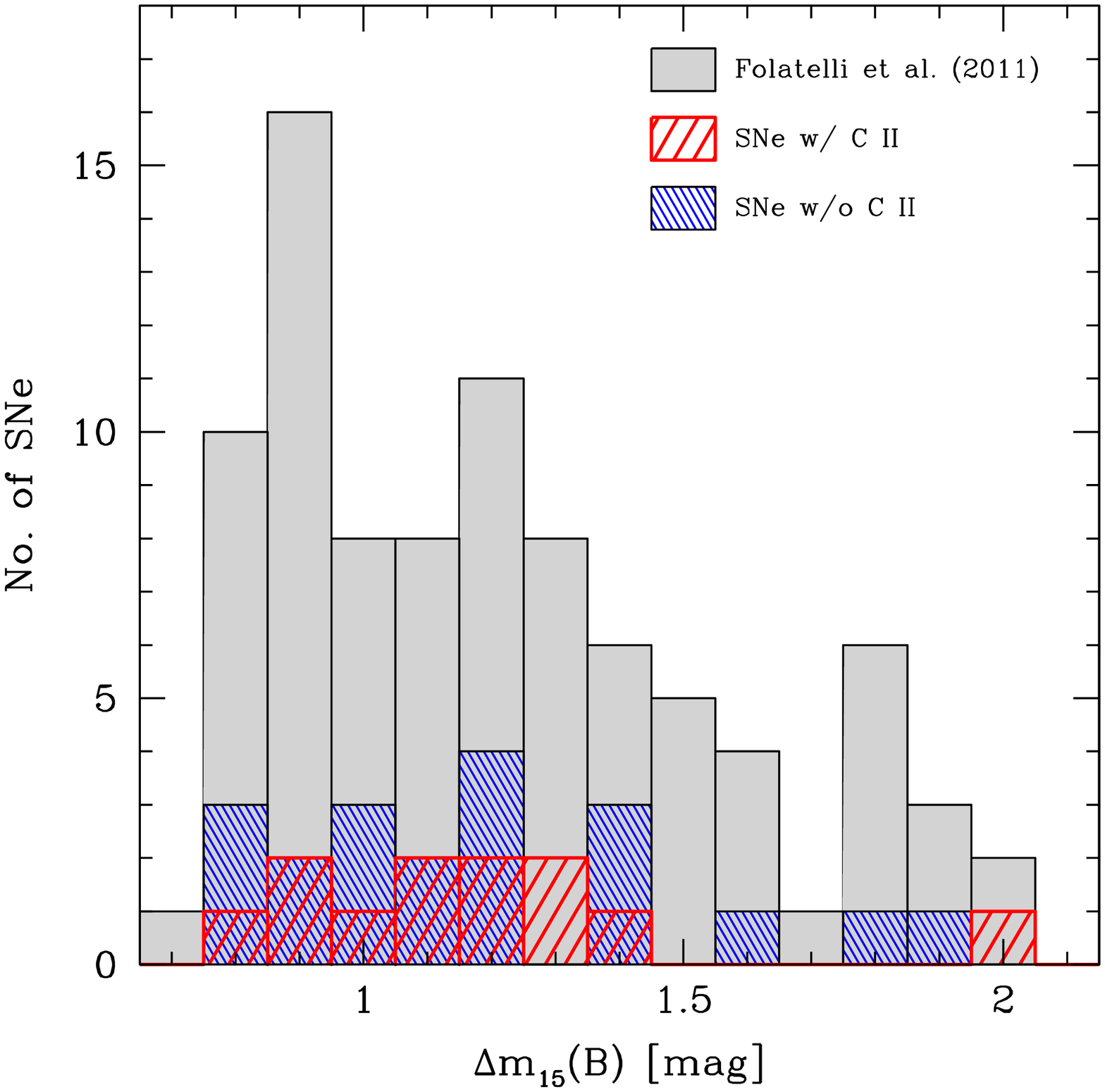}{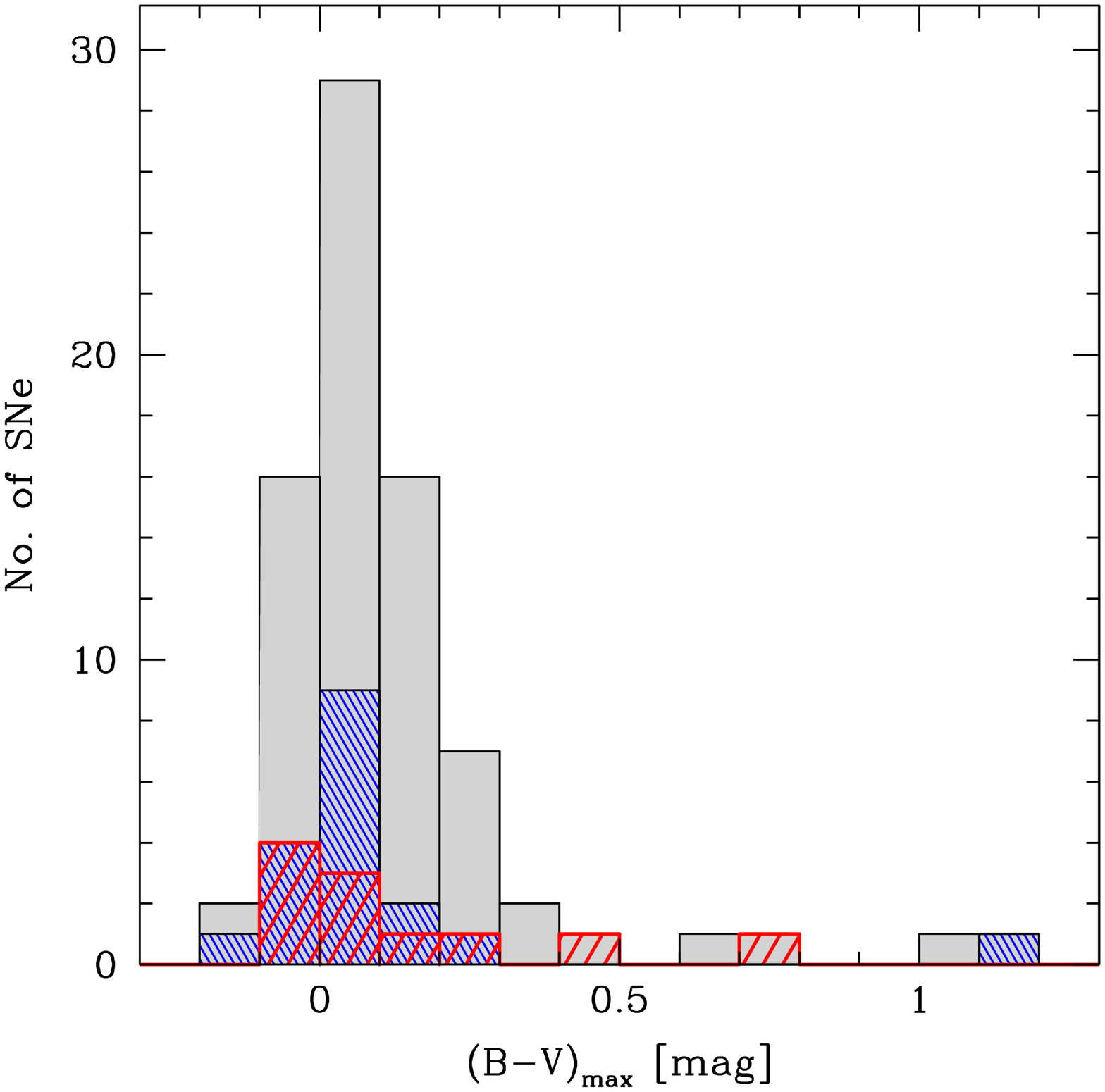}
\caption{({\em Left}) Distribution of light-curve decline rates, \dm,
  for \sneia\ with carbon ({\em black hashed}), without carbon ({\em
    light gray hashed}), and the complete sample of
  \citet{folatelli11} ({\em solid gray}). ({\em Right}) Distribution
  of observed $(B-V)$ pseudo-colors at maximum light for SNe with
  \dm\ $\leq 1.6$ mag. The shading code is the same as in the left
  panel. \label{fig:dmbvhist}}
\end{figure}

Figure~\ref{fig:dmbvhist} (left panel) shows the distribution of
light-curve decline rates, \dm\, for both groups. As a
reference, we also show the distribution of the complete sample of
\citet{folatelli11}. The group of \sneia\ with carbon covers a wide
range of decline rates, including a fast-declining (1991bg-like)
event. The average decline rate of all three samples (i.e., the
complete sample, and those with and without carbon) is $1.2 \pm 0.3$
mag. Unfortunately, the samples sizes are too small to pursue a
robust statistical test of the similarity among the distributions.

Shown in the right-hand panel of Figure~\ref{fig:dmbvhist} is the
distribution of observed $(B-V)$ pseudo-colors at maximum light for
\sneia\ with and without carbon, and for the reference sample. To
avoid confusion with the intrinsically-red CL subclass of \sneia,
we have excluded all SNe with \dm\ $> 1.65$ mag. We can
see a tendency in the distribution of SNe with carbon to peak at a bluer
color, although the statistics may be affected by the small number of
cases. As pointed out recently \citep{pignata08,wang09,foley11,maeda11}, 
\sneia\ with high expansion velocities tend to show redder colors at
maximum light. \citet{tanaka08} suggested that the underlying cause of
the difference in colors between HVG and LVG SNe is a variation in
effective temperature which arises from a difference in photospheric
velocity. The fact that SNe with carbon are all low-velocity
objects may cause the observed difference in the color distributions.

In order to further study possible differences in the intrinsic
colors of \sneia\ with and without carbon, we have selected those
object in each group which can be considered to have suffered little
reddening. Following the criteria of \citet{folatelli10b}, we picked
SNe that (a) occurred in E/S0 galaxies or that appeared outside the
nuclei and arms of spiral galaxies, and (b) which did not show any
sign of interstellar \ion{Na}{1} D absorption in their early-time
spectra. To these we added a few objects whose observed $(B-V)$ colors 
one month past maximum light were compatible with the intrinsic color
law given in Equation~(2) of \citet{folatelli10b}. The SNe in this
low-reddening sample are indicated with a ``Y'' in column 8 of
Table~\ref{tab:sne}. 

\begin{figure}[htpb]
\epsscale{1.0}
\plotone{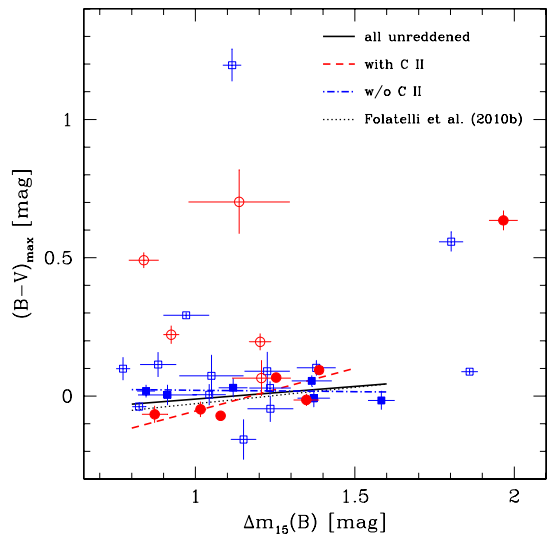}
\caption{$(B-V)$ pseudo-colors at maximum light versus \dm. Solid
  symbols indicate \sneia\ with low reddening (see text). Circles
  mark SNe with carbon and squares, SNe without carbon. Straight-line
  fits to low-reddening objects are shown with a dashed line for SNe
  with carbon, dot-dashed line for SNe without carbon, and solid line
  for both groups together. The dotted
  line is the fit given by \citet{folatelli10b} using a different
  sample. \label{fig:bvdm}} 
\end{figure}

Figure~\ref{fig:bvdm} shows the relation between color and decline rate 
for this sample of low-reddening SNe, with and without carbon 
absorption. Table~\ref{tab:bvdm} lists the resulting parameters of the
straight-line fits performed on different subsamples. We have adopted
a Markov Chain Monte Carlo (MCMC) approach to perform the fits,
with a full covariance matrix representation of the
uncertainties in $(B-V)$ and \dm. We have also added an ``intrinsic dispersion'' term,
$\sigma_I$, to the uncertainties which accounts for the extra dispersion in
the data with respect to the straight-line fit. This term is left as a
free parameter and its resulting values are given in
Table~\ref{tab:bvdm}. The ``uncertainties'' in $\sigma_I$ provide an
estimate of the precision to which the MCMC method can determine this
intrinsic dispersion. Note that once again the Cool SNe have been
excluded by requiring \dm\ $\leq 1.6$ mag.  

As shown in Table~\ref{tab:bvdm} and Figure~\ref{fig:bvdm}, the
relation between color and decline rate for the combined sample of SNe
with low-reddening is compatible with the fit given by
\citet{folatelli10b}.  On the other hand, if the 
sample is divided between those SNe with carbon and those without, a
steeper relation is found for the former than for the latter
SNe. Based on the posterior probability distributions resulting from
the MCMC analysis, we find that the intercepts and slopes of both fits
are different at the 80\% and 90\% confidence level,
respectively. We must note that all of the SNe used
in the fits ---those which are assumed to be unreddened--- happen to
belong to the LVG class, so the difference cannot be explained by
color variations between LVG and HVG SNe. Clearly, this difference
must be confirmed using a larger SN sample.

\subsection{Luminosity}
\label{sec:lum}

Now we compare the luminosities of \sneia\ with and without
carbon. For this purpose, we compute distance moduli based on the
redshifts and assuming a standard \omm--\oml\ model of the universe 
\citep[see Equation~(5) of][]{folatelli10b}, which is accurate enough for
the range of redshifts in the current sample. To avoid uncertainties
larger than about 10\% in the distances due to peculiar motions of
galaxies in the local universe, the sample is cut by imposing that $z>
0.01$.

\begin{figure}[htpb]
\epsscale{1.0}
\plottwo{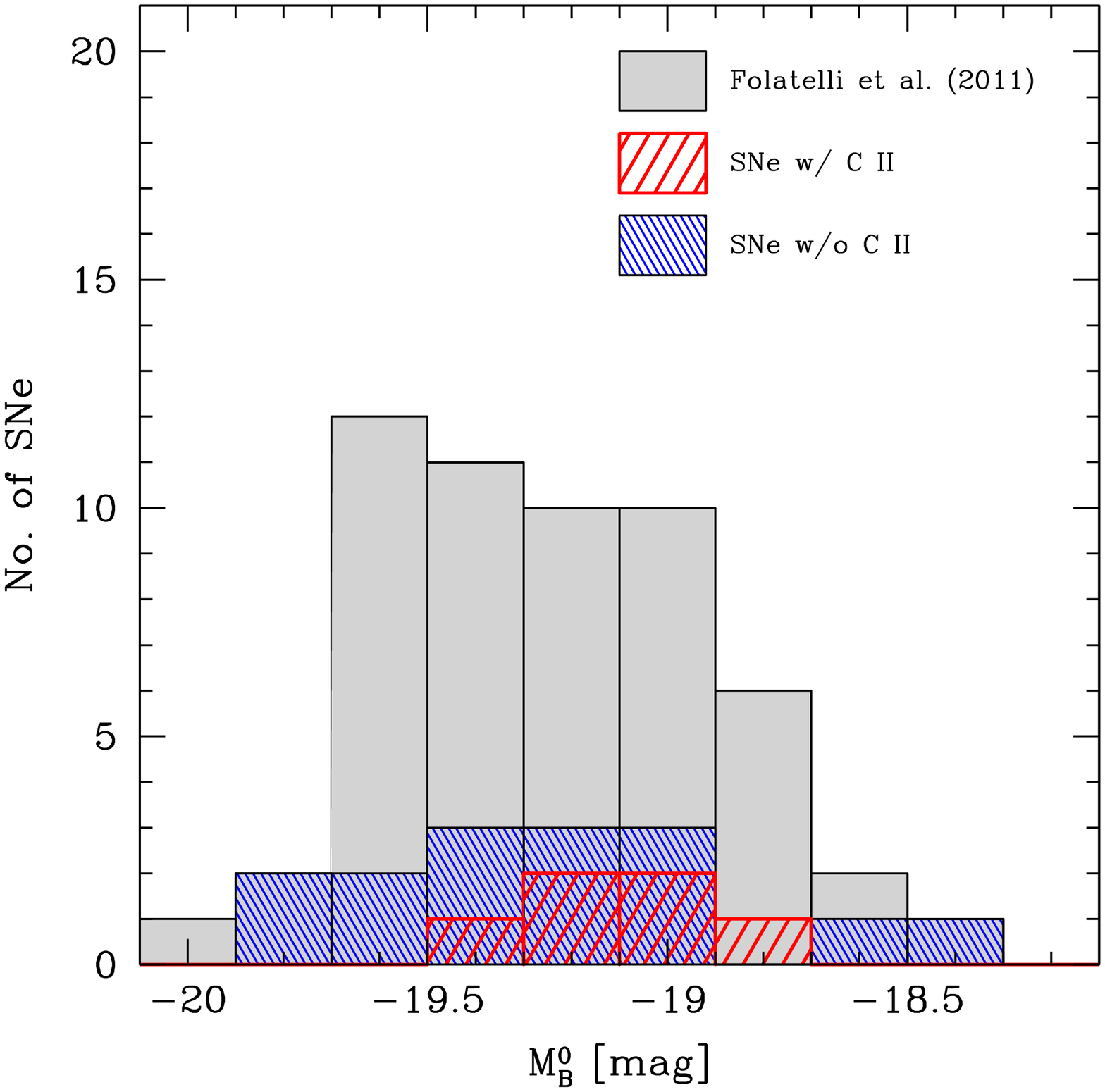}{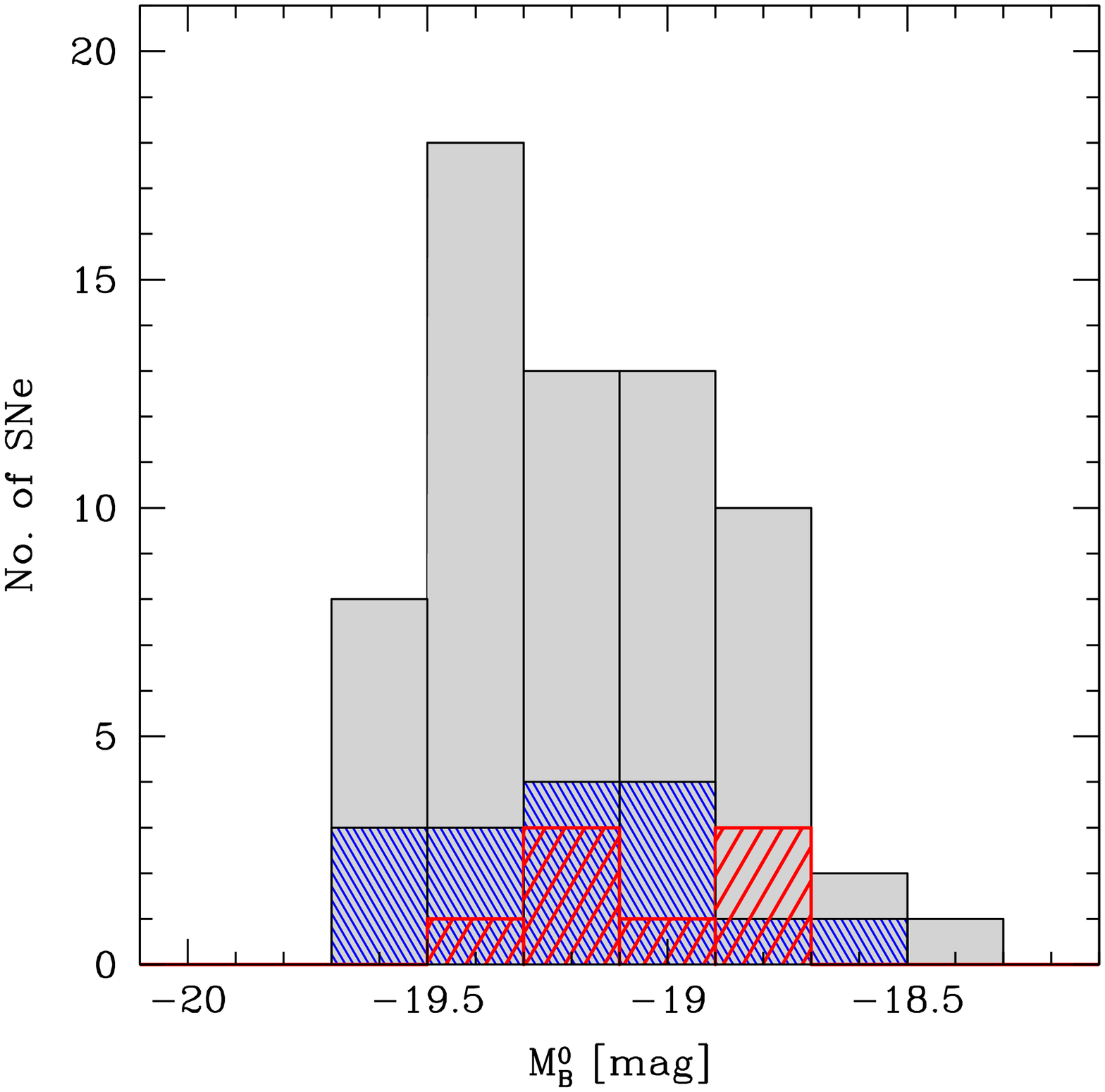}
\caption{Distributions of extinction-corrected $B$-band absolute peak
  magnitudes. ({\em Left}) Extinction corrections derived using
  $R_V=3.1$ and cutting the sample so that $(B-V)<0.2$ mag. ({\em
    Right}) Including the complete range of $(B-V)$ colors, but
  adopting a low value of $R_B=2.74$ \citep{folatelli10b}. The shading
  code is the same as in Figure~\ref{fig:dmbvhist}. \label{fig:M0bvhist}}
\end{figure}

Figure~\ref{fig:M0bvhist} shows the histograms of $B$-band absolute
peak magnitudes corrected for Milky-Way and host-galaxy extinction. In
order to derive the latter correction, color excesses $E(B-V)$ were
computed from the differences between observed $(B-V)$ pseudo-colors
at maximum light and the law of intrinsic color versus \dm\ provided
by \citet{folatelli10b}. Color excesses were converted into
extinctions by multiplying by a total-to-selective absorption
coefficient equivalent to the Galactic average of $R_V=3.1$. To avoid
large uncertainties in the extinction correction, on the left panel of
Figure~\ref{fig:M0bvhist} we cut the sample by requiring that
$(B-V)_{\mathrm{max}} < 0.2$ mag, after correcting for Milky-Way
extinction. We note an apparent difference 
in the extinction-corrected peak luminosities in the sense that \sneia\ with
carbon are on average {\em fainter} than those without, and also than
the reference sample. This difference is however statistically
marginal. The values are  $\langle M_B^0\rangle = -19.25 \pm
0.33$ mag for 55 objects in the reference sample, and $\langle
M_B^0\rangle = -19.23 \pm 0.38$ mag for 15 SNe with no carbon
detection. For the 6 SNe with carbon, this value is $\langle
M_B^0\rangle = -19.10 \pm 0.21$ mag, i.e., $0.13$ mag fainter but well
within the dispersion. 

A similar effect is seen in the histograms of the right-hand panel of
Figure~\ref{fig:M0bvhist}, where extinction-corrected, absolute peak
magnitudes are shown, this time including objects with $(B-V)_{\mathrm{max}} 
\geq 0.2$ mag, but adopting a low value of the total-to-selective
absorption coefficient of $R_B=2.74$, as derived from Hubble fits by
\citet{folatelli10b}. The differences in luminosity may in part be due
to inaccurate extinction corrections caused by a difference in
intrinsic colors as that seen in Figure~\ref{fig:bvdm}. If, as
suggested by Figure~\ref{fig:bvdm}, SNe with carbon show bluer
intrinsic colors then the extinction corrections for this group would be
underestimated. In fact, if we adopt a different extinction correction
for each group of SNe based on the intrinsic-color laws given in
Table~\ref{tab:bvdm}, the differences in average extinction-corrected peak
magnitudes nearly vanish. This may indicate that the differences
between both groups are related to differences in intrinsic colors
rather than intrinsic luminosities. 

\begin{figure}[htpb]
\epsscale{1.0}
\plotone{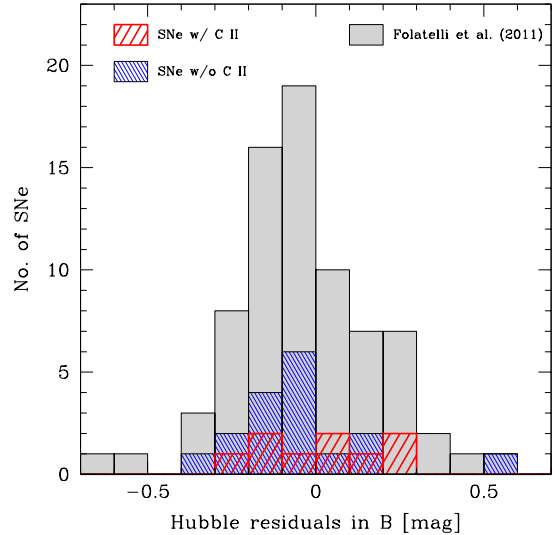}
\caption{Distribution of $B$-band Hubble residuals based on Fit 2 of
  Table~8 in \citet{folatelli10b}. Shading code is the same as in
  Figure~\ref{fig:dmbvhist}. \label{fig:reshist}} 
\end{figure}

In Figure~\ref{fig:reshist} we show the distributions of Hubble
residuals in $B$ band, $\Delta M_B$. For this purpose, we adopted the parameters
of fit 2 in Table 8 of \citet{folatelli10b} which correspond to a fit of
distance moduli versus \dm\ and $(B-V)$ pseudo-colors. The discrepancy
between SNe with and without carbon in this case is seen in the
same direction as above, although reduced to $0.08$ mag. The mean
residuals and their standard deviations are $\langle 
\Delta M_B \rangle = 0.03 \pm 0.18$ mag and $\langle \Delta M_B
\rangle = -0.05 \pm 0.21$ mag for \sneia\ with and without carbon,
respectively. 

\begin{figure}[htpb]
\epsscale{1.0}
\plotone{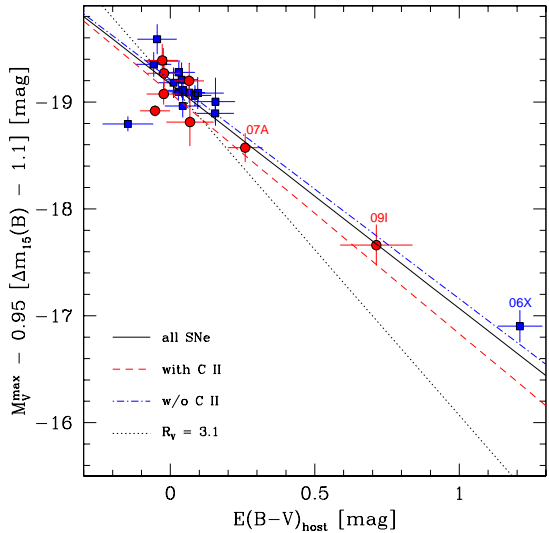}
\caption{Decline-rate corrected $V$-band absolute peak magnitude
  versus $E(B-V)$ color excess for \sneia\ with ({\em circles}) and
  without ({\em squares}) carbon. The solid line is a straight-line
  fit to all the data points, while the dashed and dot-dashed lines
  correspond to the \sneia\ with and without carbon, respectively. As
  a reference, the dotted line corresponds to the case of
  $R_V=3.1$. All three fits are in agreement with each other and
  favor a low value of $R_V$. Three SNe with the largest reddening
  ---those which weigh most in the fits--- are
  indicated.\label{fig:Mvce}}  
\end{figure}

Finally, we consider the effect of reddening on absolute peak
magnitudes. For this purpose, we take the values of $M_V$ at maximum
light corrected for decline rate by adopting the slope of fit 7 in
Table 9 of \citet{folatelli10b}. In Figure~\ref{fig:Mvce} we show these
corrected peak magnitudes versus $E(B-V)$ color excesses derived as
explained above. With the current sample, we find no evidence of
different extinction properties between \sneia\ with and without
carbon. The data favors a low value of $R_V \approx 2.1$ as has been
found in previous studies of \sneia\ data.

\section{DISCUSSION \& CONCLUSIONS}
\label{sec:dis}

The most striking result of this paper is the large incidence of carbon 
lines in \sneia. In the CSP sample, we have found evidence of \ion{C}{2} 
$\lambda$6580 absorption in at least 30\% of the objects, which agrees
closely with the findings of \citet{parrent11} based on a completely 
independent sample of publicly-available spectra. The
  identification of this absorption with other elements, such as hydrogen,
  is considered unlikely (see Section~\ref{sec:H?}). In addition, we have 
shown that nearly 10\% of the \sneia\ in the CSP sample show a suppressed 
emission component of the \ion{Si}{2} $\lambda$6355 line, which SYNOW models 
indicate may be caused by absorption due to the same \ion{C}{2} $\lambda$6580 
line. Taken together, these results imply that nearly half of all \sneia\ 
show carbon in their spectra at early phases, a finding that contradicts 
previous conclusions \citep[e.g., see][]{thomas07,branch07}.  The CSP 
spectra show clearly that the detection of \ion{C}{2} absorption is a strong
function of the age, falling from $>$40\% at 10 days before $B$ maximum
to zero by the time of maximum itself. These results almost certainly
represent a lower limit on the presence of unburned material in \sneia\
since very few events have been observed at even earlier phases.  Moreover,
low signal-to-noise observations, or line blending when expansion 
velocities are high, also contribute to the difficulty of detecting the
\ion{C}{2} $\lambda$6580 line.  As we have shown, the latter effect can
explain at least in part the absence of detectable carbon absorption
in members of the high  
velocity gradient \citep{benetti05} or BL \citep{branch06} subclasses found by
\citet{parrent11} and ourselves. As additional data are obtained
  at epochs of $-10$ days or earlier, it will be possible to corroborate 
whether, as we suspect, the majority of \sneia\ show carbon at some 
stage of their evolution.  The theoretical implications of such findings
are potentially large, in particular as pertains to the mechanism by which 
the explosive flame propagates inside the WD star.

The second remarkable finding from our study of the CSP spectra and the
work of \citet{parrent11} is that, assuming that the absorption observed at
about 6300 \AA\ is due to \ion{C}{2} $\lambda$6580, then carbon must be
present in very deep regions, corresponding to velocities as low as $v$
$\approx$ 11000 km s$^{-1}$, which is right above the bulk of silicon
\citep{mazzali07}. This is well below the expected limit imposed by
one-dimensional models, and points directly
to large mixing effects and/or possible departures from spherical
symmetry or clumpiness. 

In spherically symmetric models, irrespective of the details of the flame 
propagation (deflagration or detonation), the production of a large amount of 
$^{56}$Ni as typically observed \citep[$\sim 0.6 \, M_\odot$;
  see][]{stritzinger06} requires that 
the strength of the flame ---either its speed or the detonation
transition density--- is high, which must lead to the total
consumption of carbon below 15000 km s$^{-1}$
\citep[e.g.][]{nomoto84,iwamoto99}, and even below 20000 km s$^{-1}$ in
detonation models \citep[e.g.][]{shigeyama92}.
The situation is different for non-spherical models. For example,
the off-center ignition model of \citet{maeda10b} \citep[see
  also][]{kasen09} synthesizes $0.54 \, M_\odot$ of $^{56}$Ni, despite
the relatively low density at which the detonation is  
triggered. The low transition density in their model results in less
burning in the outer layers, and thereby the existence of carbon
with a mass fraction of $\sim 0.1$ at velocities as low as
$\sim$13000 km s$^{-1}$. This is still larger than the observed
velocity, but suggests that the non-spherical effects may be important in
understanding the detection of carbon deep in the ejecta. This argument
is based on the effect of a global asymmetry produced by the off-center
ignition. Similar conclusions may be reached from possible effects of
mixing or clump formation in multi-dimensional models. 

The observed rate of carbon detections could
thus provide information on the distribution of C-rich material, and
the amount and size of the clumps. Unfortunately, the current
uncertainties in the detection rate complicate these geometrical
considerations. Nevertheless, if carbon were present in a single clump
or asymmetric blob, the tight dispersion in expansion velocities of
\ion{C}{2} $\lambda$6580 compared with silicon observed in
Figure~\ref{fig:Cvel} would be surprising \citep[see
  also][]{parrent11}. 

In view of recent 
results which connect asymmetries in the ejecta with the observed 
spectroscopic and photometric diversity \citep{maund10,maeda11}, it is worth 
studying possible connections with the presence of unburned material in the 
ejecta. Breaking our sample of CSP \sneia\ into two groups depending on
the presence or absence of high-velocity components of \ion{Ca}{2} 
absorption ---a possible indicator of asymmetry \citep{tanaka06}--- we
find no significant distinction in the frequency of detectable carbon.
There is an indication in our data that \sneia\ with carbon show slightly 
bluer colors and lower luminosities at maximum light when compared with the
rest of the sample. These are interesting hints of a possible origin
of the observed diversity arising from incomplete burning which may be
related to the asymmetry scenario of \citet{maeda11}, but the statistical
significance of these results is limited by the sample size.

As shown in Section~\ref{sec:MC}, modeling indicates that low masses of
carbon in the range $M(C) \sim 10^{-3}$ -- $10^{-2}\, M_\odot$ are
enough to reproduce the observed spectral features, although
high-velocity components, undetectable due to blending with the
\ion{Si}{2} $\lambda$6355 line, could increase this value by as much
as an order of magnitude \citep{tanaka11}. Such small 
amounts of unburned material would make it difficult to detect other possible
signatures in the observations. For instance, the non-detection of \ion{C}{1}
features in near-infrared spectra reported by \citet{marion06} could
be explained by abundance and ionization state considerations
\citep{tanaka08}. 

Another interesting possibility is an enhancement of oxygen versus
magnesium in SNe which suffered less complete burning. While
  oxygen may in part be primordial, magnesium is only a 
  product of carbon burning. Determining the distribution of these
  elements in the outer ejecta may shed light on the evolution of the
  burning process \citep{mazzali08}. This could be attained with a
  temporal followup of the dominant spectral lines of these species,
  which at early times are those of \ion{O}{1} and
  \ion{Mg}{2}. Unfortunately, the study of such lines is complicated
  from an observational viewpoint. The main \ion{O}{1} line at
optical wavelengths usually coincides with the 
telluric A band, which makes it difficult to obtain an accurate
measure of its strength from the ground. Other \ion{O}{1} and
\ion{Mg}{2} lines appear blended with each other or with lines of
other elements (e.g, see Figure~\ref{fig:Csyn05el}). In this sense,
the study of near-IR spectra may offer the possibility of finding
isolated oxygen and magnesium lines \citep{marion09}.

Figure~\ref{fig:Cfrac} clearly shows that the
definitive assessment of the incidence of unburned material in
\sneia\ require a survey of early-time spectra ---obtained at least
ten days before maximum light--- covering the region of the \ion{C}{2}
lines. High signal-to-noise data are required in order to explore
the possible presence of carbon in high-velocity \sneia, in
particular by observing the relatively unblended but weak \ion{C}{2}
$\lambda$7234 line. 

\acknowledgments 
We are grateful to Wojtek Krzeminski for his dedicated efforts during
the CSP campaigns. We would also like to thank Luc Dessart for his
useful suggestions, and 
Joseph Anderson, Francisco F\"orster, Giuliano Pignata, and the rest of
the MCSS team for interesting discussions on the topic of this
paper. This material is based upon work supported by NSF under  
grants AST--0306969, AST-0908886, AST--0607438, and AST-1008343. 
The Dark Cosmology Centre is funded by the Danish NSF. This research
is supported by the World Premier International Research Center
Initiative (WPI Initiative), MEXT, Japan. G.F. acknowledges financial
support by Grant-in-Aid for Scientific Research for Young Scientists
(23740175). M.H. acknowledges support by CONICYT through grants
FONDECYT Regular 1060808, Centro de Astrofisica FONDAP 15010003,
Centro BASAL CATA (PFB--06), and by the Millennium Center for
Supernova Science (P06--045-F).


\clearpage
\begin{deluxetable}{lcccccccccc} 
\tabletypesize{\scriptsize} 
\tablecolumns{11} 
\tablewidth{0pt} 
\tablecaption{Sample of SNe~Ia\label{tab:sne}} 
\tablehead{ 
\colhead{} & \colhead{First} & \colhead{\ion{C}{2}} & \colhead{$m_B$} & \colhead{$(B-V)_{\mathrm{max}}$} & \colhead{\dm} & \colhead{$E(B-V)_{\mathrm{host}}$} & \colhead{Low} & \colhead{} & \colhead{$M_B$} & \colhead{$\Delta M_B$} \\ 
\colhead{SN} & \colhead{Epoch} & \colhead{Class} & \colhead{(mag)} & \colhead{(mag)} & \colhead{(mag)} & \colhead{(mag)} & \colhead{Redd.} & \colhead{$z_{\mathrm{Helio}}$} & \colhead{(mag)} & \colhead{(mag)} \\ 
\colhead{(1)} & \colhead{(2)} & \colhead{(3)} & \colhead{(4)} & \colhead{(5)} & \colhead{(6)} & \colhead{(7)} & \colhead{(8)} & \colhead{(9)} & \colhead{(10)} & \colhead{(11)} 
} 
\startdata
2004dt & $  -8.7$ &        N & $ 15.000$($035$) & $ -0.046$($047$) & $  1.24$($07$) & $ -0.046$($068$) &  \nodata & $ 0.01973$ & $ -19.51$($12$) & $  -0.39$($18$) \\ 
2004ef & $  -8.5$ &        N & $ 16.828$($019$) & $  0.102$($027$) & $  1.38$($06$) & $  0.085$($057$) &  \nodata & $ 0.03099$ & $ -18.69$($08$) & $  -0.09$($12$) \\ 
2004eo & $ -11.1$ &        N & $ 15.067$($011$) & $  0.055$($020$) & $  1.36$($06$) & $  0.039$($054$) &        Y & $ 0.01570$ & $ -18.90$($15$) & $  -0.16$($17$) \\ 
2004ey & $  -0.8$ &        N & $ 14.713$($011$) & $ -0.063$($022$) & $  0.93$($01$) & $ -0.027$($055$) &  \nodata & $ 0.01579$ & $ -19.24$($15$) & $   0.15$($16$) \\ 
2004gs & $  -3.3$ &        N & $ 17.132$($017$) & $  0.158$($044$) & $  1.61$($08$) & \nodata &  \nodata & $ 0.02665$ & $ -18.21$($08$) & $   0.07$($16$) \\ 
2004gu & $  -2.9$ &        N & $ 17.478$($017$) & $  0.189$($021$) & $  0.89$($30$) & $  0.231$($054$) &  \nodata & $ 0.04586$ & $ -19.05$($05$) & $  -0.33$($23$) \\ 
2005M  & $  -0.1$ &        N & $ 15.924$($038$) & $  0.018$($039$) & $  0.84$($07$) & $  0.065$($063$) &  \nodata & $ 0.02201$ & $ -19.02$($10$) & $   0.21$($16$) \\ 
2005ag & $   0.3$ &        N & $ 18.447$($060$) & $ -0.003$($125$) & $  0.92$($17$) & $  0.035$($135$) &  \nodata & $ 0.07943$ & $ -19.30$($07$) & $  -0.07$($37$) \\ 
2005bl & $  -8.4$ &       F? & $ 18.221$($031$) & $  0.558$($035$) & $  1.80$($04$) & \nodata &  \nodata & $ 0.02406$ & $ -16.92$($09$) & $   0.12$($14$) \\ 
2005bo & $  -1.0$ &        N & $ 15.662$($009$) & $  0.290$($017$) & $  1.30$($08$) & $  0.282$($053$) &  \nodata & $ 0.01390$ & $ -18.34$($15$) & $  -0.19$($16$) \\ 
2005el & $  -7.3$ &        A & $ 14.821$($021$) & $ -0.014$($022$) & $  1.35$($04$) & $ -0.028$($055$) &        Y & $ 0.01491$ & $ -19.17$($15$) & $  -0.22$($16$) \\ 
2005eq & $  -5.2$ &        N & $ 16.306$($026$) & $  0.099$($042$) & $  0.77$($02$) & $  0.155$($065$) &  \nodata & $ 0.02898$ & $ -19.11$($08$) & $  -0.05$($14$) \\ 
2005hc & $  -5.2$ &        N & $ 17.305$($033$) & $  0.004$($036$) & $  0.91$($09$) & $  0.043$($062$) &        Y & $ 0.04594$ & $ -19.14$($06$) & $   0.08$($13$) \\ 
2005ke & $  -1.4$ &        N & $ 14.772$($021$) & $  0.661$($023$) & $  1.76$($05$) & \nodata &  \nodata & $ 0.00488$ & $ -17.07$($08$) & $  -0.28$($11$) \\ 
2005ki & $  -7.6$ &        N & $ 15.543$($031$) & $ -0.008$($031$) & $  1.37$($05$) & $ -0.025$($059$) &        Y & $ 0.01921$ & $ -19.13$($11$) & $  -0.22$($15$) \\ 
2006D  & $  -6.1$ &        A & $ 14.136$($006$) & $  0.094$($013$) & $  1.39$($01$) & $  0.075$($052$) &        Y & $ 0.00853$ & \nodata & \nodata \\ 
2006X  & $ -10.4$ &        N & $ 15.218$($048$) & $  1.196$($058$) & $  1.11$($03$) & $  1.210$($077$) &  \nodata & $ 0.00524$ & $ -15.69$($15$) & $   0.11$($22$) \\ 
2006ax & $ -10.5$ &        F & $ 15.002$($028$) & $ -0.048$($028$) & $  1.02$($02$) & $ -0.022$($057$) &        Y & $ 0.01674$ & $ -19.40$($13$) & $  -0.11$($15$) \\ 
2006dd & $ -10.8$ &        A & $ 12.241$($009$) & $ -0.071$($013$) & $  1.08$($01$) & $ -0.053$($052$) &        Y & $ 0.00587$ & $ -19.01$($05$) & $   0.29$($06$) \\ 
2006gt & $  -0.3$ &        N & $ 18.229$($020$) & $  0.256$($024$) & $  1.85$($29$) & \nodata &  \nodata & $ 0.04477$ & $ -18.15$($06$) & $  -0.31$($23$) \\ 
2006hx & $  -9.6$ &        N & $ 17.510$($031$) & $ -0.157$($073$) & $  1.15$($04$) & $ -0.147$($088$) &  \nodata & $ 0.04549$ & $ -18.90$($06$) & $   0.58$($21$) \\ 
2006kf & $  -5.1$ &        N & $ 15.817$($005$) & $ -0.016$($033$) & $  1.58$($04$) & $ -0.058$($060$) &        Y & $ 0.02130$ & $ -18.91$($11$) & $  -0.13$($14$) \\ 
2006mr & $  -2.3$ &        N & $ 15.345$($011$) & $  0.708$($121$) & $  1.78$($04$) & \nodata &  \nodata & $ 0.00587$ & $ -15.90$($05$) & $   0.75$($34$) \\ 
2006os & $  -1.9$ &        N & $ 17.500$($020$) & $  0.314$($024$) & $  1.44$($07$) & $  0.289$($055$) &  \nodata & $ 0.03281$ & $ -18.19$($07$) & $  -0.21$($11$) \\ 
2007A  & $  -6.2$ &        F & $ 15.695$($012$) & $  0.222$($033$) & $  0.92$($02$) & $  0.259$($060$) &  \nodata & $ 0.01765$ & $ -18.52$($13$) & $   0.09$($16$) \\ 
2007S  & $ -11.9$ &  \nodata & $ 15.789$($013$) & $  0.371$($031$) & $  0.95$($03$) & $  0.405$($059$) &  \nodata & $ 0.01388$ & $ -18.22$($15$) & $  -0.03$($17$) \\ 
2007af & $ -11.1$ &        A & $ 13.330$($014$) & $  0.196$($030$) & $  1.20$($03$) & $  0.200$($059$) &  \nodata & $ 0.00546$ & \nodata & \nodata \\ 
2007ai & $  -2.7$ &        N & $ 16.805$($009$) & $  0.119$($011$) & $  0.74$($01$) & $  0.178$($051$) &  \nodata & $ 0.03166$ & $ -18.87$($07$) & $   0.16$($08$) \\ 
2007as & $  -2.2$ &        N & $ 15.387$($006$) & $  0.052$($008$) & $  1.22$($01$) & $  0.054$($051$) &  \nodata & $ 0.01757$ & $ -19.01$($12$) & $  -0.14$($12$) \\ 
2007ax & $  -2.1$ &        N & $ 16.201$($029$) & $  0.684$($031$) & $  1.96$($09$) & \nodata &  \nodata & $ 0.00686$ & \nodata & \nodata \\ 
2007bd & $  -7.5$ &        N & $ 16.551$($022$) & $  0.029$($024$) & $  1.23$($06$) & $  0.029$($056$) &  \nodata & $ 0.03102$ & $ -19.12$($07$) & $  -0.21$($11$) \\ 
2007jg & $  -0.2$ &        N & $ 17.181$($045$) & $  0.028$($048$) & $  1.31$($08$) & $  0.019$($070$) &  \nodata & $ 0.03713$ & $ -18.80$($08$) & $   0.06$($16$) \\ 
2007le & $  -9.2$ &       F? & $ 13.875$($009$) & $  0.292$($011$) & $  0.97$($07$) & $  0.323$($051$) &  \nodata & $ 0.00672$ & \nodata & \nodata \\ 
2007on & $  -3.9$ &       F? & $ 13.023$($008$) & $  0.088$($011$) & $  1.86$($03$) & \nodata &        Y & $ 0.00649$ & \nodata & \nodata \\ 
2008ar & $  -8.1$ &  \nodata & $ 16.259$($038$) & $  0.100$($054$) & $  0.88$($07$) & $  0.142$($073$) &  \nodata & $ 0.02615$ & $ -19.06$($09$) & $  -0.09$($18$) \\ 
2008bc & $ -10.1$ &        N & $ 14.624$($007$) & $ -0.038$($015$) & $  0.82$($02$) & $  0.011$($052$) &  \nodata & $ 0.01509$ & $ -19.48$($14$) & $  -0.08$($15$) \\ 
2008bf & $  -8.5$ &        A & $ 15.779$($022$) & $ -0.066$($031$) & $  0.87$($04$) & $ -0.023$($059$) &        Y & $ 0.02403$ & $ -19.36$($09$) & $   0.08$($13$) \\ 
2008bq & $   0.3$ &        N & $ 16.661$($006$) & $  0.052$($008$) & $  0.92$($01$) & $  0.090$($051$) &  \nodata & $ 0.03400$ & $ -19.18$($06$) & $  -0.10$($07$) \\ 
2008fp & $  -3.7$ &        A & $ 13.829$($020$) & $  0.491$($028$) & $  0.84$($05$) & $  0.538$($057$) &  \nodata & $ 0.00566$ & \nodata & \nodata \\ 
2008gl & $  -1.5$ &        N & $ 16.845$($022$) & $  0.050$($031$) & $  1.29$($04$) & $  0.043$($059$) &  \nodata & $ 0.03402$ & $ -18.90$($07$) & $  -0.09$($12$) \\ 
2008gp & $  -6.0$ &  \nodata & $ 16.455$($031$) & $  0.033$($043$) & $  0.97$($11$) & $  0.065$($066$) &  \nodata & $ 0.03341$ & $ -19.28$($08$) & $  -0.19$($16$) \\ 
2008hj & $  -6.6$ &        N & $ 16.844$($027$) & $  0.005$($038$) & $  1.04$($05$) & $  0.028$($063$) &  \nodata & $ 0.03789$ & $ -19.14$($07$) & $  -0.03$($13$) \\ 
2008hv & $  -6.6$ &        A & $ 14.799$($006$) & $  0.067$($009$) & $  1.25$($01$) & $  0.065$($051$) &        Y & $ 0.01255$ & $ -18.99$($16$) & $  -0.19$($16$) \\ 
2009D  & $  -3.9$ &       N? & $ 15.789$($016$) & $  0.018$($023$) & $  0.84$($03$) & $  0.065$($055$) &        Y & $ 0.02501$ & $ -19.31$($09$) & $  -0.08$($11$) \\ 
2009F  & $  -5.4$ &        A & $ 16.929$($025$) & $  0.635$($036$) & $  1.97$($04$) & \nodata &        Y & $ 0.01296$ & $ -16.75$($17$) & $  -0.04$($20$) \\ 
2009I  & $  -5.1$ &        F & $ 18.247$($081$) & $  0.702$($115$) & $  1.14$($16$) & $  0.713$($125$) &  \nodata & $ 0.02617$ & $ -16.92$($12$) & $   0.22$($36$) \\ 
2009Y  & $  -5.5$ &       N? & $ 14.042$($032$) & $  0.114$($045$) & $  0.88$($06$) & $  0.156$($067$) &  \nodata & $ 0.00935$ & $ -19.09$($22$) & $  -0.16$($26$) \\ 
2009aa & $  -7.8$ &       F? & $ 16.355$($023$) & $  0.030$($033$) & $  1.12$($04$) & $  0.044$($060$) &        Y & $ 0.02731$ & $ -19.06$($08$) & $  -0.07$($12$) \\ 
2009ab & $ -10.2$ &        A & $ 14.652$($046$) & $  0.065$($065$) & $  1.21$($09$) & $  0.068$($082$) &  \nodata & $ 0.01117$ & $ -18.65$($21$) & $   0.19$($28$) \\ 
2009ad & $  -8.3$ &  \nodata & $ 16.074$($096$) & $ -0.006$($136$) & $  0.88$($17$) & $  0.036$($145$) &  \nodata & $ 0.02840$ & $ -19.33$($12$) & $  -0.07$($41$) \\ 
2009ag & $  -3.5$ &       F? & $ 14.589$($041$) & $  0.241$($058$) & $  0.97$($08$) & $  0.273$($077$) &  \nodata & $ 0.00864$ & \nodata & \nodata \\ 
2009cz & $  -4.1$ &        N & $ 15.786$($053$) & $  0.073$($075$) & $  1.05$($10$) & $  0.095$($090$) &  \nodata & $ 0.02114$ & $ -19.06$($11$) & $  -0.13$($25$) \\ 
2009dc & $  -7.7$ &       SC & $ 15.141$($066$) & $  0.145$($093$) & $  0.59$($11$) & $  0.222$($106$) &  \nodata & $ 0.02139$ & $ -19.68$($12$) & $  -0.61$($29$) \\ 
\enddata 
\tablecomments{Columns: (1) SN name; (2) Epoch of the first spectroscopic observation relative to $B$-band maximum light; (3) \ion{C}{2} detection type (see Section~\ref{sec:CIIse}): Absorption (A); Flat emission (F); No detection (N); (4) MW-extinction corrected, $K$-corrected, apparent $B$-band peak magnitude (uncertainties in thousandth of mag); (5) Observed pseudocolor at maximum light, corrected for MW reddening (uncertainties in thousandth of mag); (6) Observed decline rate, \dm\ (uncertainties in hundredth of mag); (7) Host galaxy color excess (uncertainties in thousandth of mag); (8) SNe with "Y" belong to the low-reddening sample; (9) Heliocentric redshift from the NASA/IPAC Extragalactic Database (NED); (10) $B$-band absolute peak magnitude (uncertainties in hundredth of mag); (11) $B$-band Hubble residual (see Section~\ref{sec:phprop}; uncertainties in hundredth of mag).}

\end{deluxetable}

\clearpage
\LongTables
\begin{deluxetable}{lccccccc}
\tabletypesize{\scriptsize}
\tablecolumns{8}
\tablewidth{0pt}
\tablecaption{List of pre-maximum spectra \label{tab:spec}}
\tablehead{
\colhead{UT Date} & \colhead{MJD} &\colhead{Phase} & \colhead{Wavelength} & \colhead{Resol.} & \colhead{\ion{C}{2}} & \colhead{$pW$(6300)} & \colhead{Ref.} \\
\colhead{} & \colhead{} & \colhead{} & \colhead{Range (\AA)} & \colhead{(\AA)} & \colhead{Class} & \colhead{(\AA)} & \colhead{} \\
\colhead{(1)} & \colhead{(2)} & \colhead{(3)} & \colhead{(4)} & \colhead{(5)} & \colhead{(6)} & \colhead{(7)} & \colhead{(8)} 
}
\startdata
\multicolumn{8}{c}{{\bf SN 2004dt}}\\
2004-08-12          & $  53230.22$ & $  -8.7$ & $   4320 -  10315$ & \nodata  & N  & $< 1.11$         & 2  \\
2004-08-12          & $  53230.67$ & $  -8.2$ & $   3140 -   8589$ & \nodata  & N  & $< 0.75$         & 2  \\
2004-08-13          & $  53231.23$ & $  -7.7$ & $   3946 -   9702$ & \nodata  & N  & $< 1.02$         & 2  \\
2004-08-13          & $  53231.63$ & $  -7.3$ & $   3170 -  10250$ & \nodata  & N  & $< 0.66$         & 2  \\
2004-08-15          & $  53233.64$ & $  -5.3$ & $   3142 -  10400$ & \nodata  & N  & $< 0.33$         & 2  \\
2004-08-15          & $  53233.69$ & $  -5.3$ & $   3617 -   8879$ & \nodata  & N  & $< 0.39$         & 2  \\
2004-08-16          & $  53233.71$ & $  -5.2$ & $   3087 -  10211$ & \nodata  & N  & $< 0.24$         & 2  \\
2004-08-16          & $  53234.57$ & $  -4.4$ & $   3236 -  10678$ & \nodata  & N  & $< 0.45$         & 2  \\
2004-08-18          & $  53235.69$ & $  -3.3$ & $   3334 -   8878$ & \nodata  & N  & \nodata          & 2  \\
2004-08-18          & $  53236.54$ & $  -2.5$ & $   3217 -  10144$ & \nodata  & N  & \nodata          & 2  \\
2004-08-19          & $  53237.18$ & $  -1.8$ & $   3317 -   8894$ & \nodata  & N  & \nodata          & 2  \\
2004-08-20          & $  53238.22$ & $  -0.8$ & $   3318 -   8822$ & \nodata  & N  & \nodata          & 2  \\
2004-08-21          & $  53239.23$ & $   0.2$ & $   3337 -   8822$ & \nodata  & N  & \nodata          & 2  \\
\multicolumn{8}{c}{{\bf SN 2004ef}}\\
2004-09-07$\dagger$ & $  53255.66$ & $  -8.5$ & $   3780 -   7323$ & $   5.0$ & N  & $< 2.01$         & 1  \\
2004-09-08          & $  53256.68$ & $  -7.5$ & $   4000 -   7183$ & $   5.0$ & N  & $< 0.99$         & 1  \\
2004-09-09$\dagger$ & $  53257.77$ & $  -6.5$ & $   3650 -   6850$ & $  14.0$ & N  & $< 1.05$         & 1  \\
2004-09-10          & $  53258.68$ & $  -5.6$ & $   4000 -   7114$ & $   5.0$ & N  & $< 8.73$         & 1  \\
2004-09-14          & $  53262.61$ & $  -1.8$ & $   3800 -   9235$ & $   8.0$ & N  & \nodata          & 1  \\
\multicolumn{8}{c}{{\bf SN 2004eo}}\\
2004-09-19          & $  53267.52$ & $ -11.1$ & $   4530 -   9618$ & $   8.0$ & N  & $< 1.47$         & 1  \\
2004-09-19          & $  53268.05$ & $ -10.6$ & $   3501 -   9204$ & $   0.8$ & N  & $< 0.30$         & 3  \\
2004-09-20          & $  53268.51$ & $ -10.1$ & $   3800 -   9235$ & $   8.0$ & N  & $< 0.42$         & 1  \\
2004-09-24          & $  53272.74$ & $  -5.9$ & $   3341 -  10700$ & $   0.2$ & N  & $< 0.21$         & 3  \\
2004-09-27          & $  53276.39$ & $  -2.3$ & $   3067 -  10676$ & $   0.6$ & N  & \nodata          & 3  \\
\multicolumn{8}{c}{{\bf SN 2004ey}}\\
2004-10-25          & $  53303.55$ & $  -0.8$ & $   3600 -   9000$ & $  14.0$ & N  & \nodata          & 1  \\
\multicolumn{8}{c}{{\bf SN 2004gs}}\\
2004-12-13          & $  53352.76$ & $  -3.3$ & $   3800 -   9235$ & $   8.0$ & N  & \nodata          & 1  \\
\multicolumn{8}{c}{{\bf SN 2004gu}}\\
2004-12-19          & $  53358.83$ & $  -2.9$ & $   3600 -   9000$ & $  14.0$ & N  & \nodata          & 1  \\
\multicolumn{8}{c}{{\bf SN 2005M}}\\
2005-02-04          & $  53405.71$ & $  -0.1$ & $   3800 -   9235$ & $   8.0$ & N  & \nodata          & 1  \\
\multicolumn{8}{c}{{\bf SN 2005W}}\\
2005-02-03          & $  53404.51$ & $  -7.5$ & $   5760 -   7380$ & $   5.0$ & F? & $< 2.52$         & 7  \\
\multicolumn{8}{c}{{\bf SN 2005ag}}\\
2005-02-12          & $  53413.89$ & $   0.3$ & $   3800 -   9235$ & $   8.0$ & N? & \nodata          & 1  \\
\multicolumn{8}{c}{{\bf SN 2005bl}}\\
2005-04-16          & $  53475.64$ & $  -8.4$ & $   3800 -   9235$ & $   8.0$ & F? & $< 2.70$         & 1  \\
2005-04-17          & $  53477.60$ & $  -6.5$ & $   3450 -   8803$ & $  10.0$ & F? & $< 4.98$         & 4  \\
2005-04-19          & $  53479.63$ & $  -4.5$ & $   3800 -   9235$ & $   8.0$ & F? & $< 5.97$         & 1  \\
2005-04-19          & $  53479.70$ & $  -4.4$ & $   4960 -  10449$ & $  11.0$ & F? & $< 0.87$         & 4  \\
\multicolumn{8}{c}{{\bf SN 2005bo}}\\
2005-04-18          & $  53478.73$ & $  -1.0$ & $   3800 -   9235$ & $   8.0$ & N  & \nodata          & 1  \\
2005-04-19          & $  53479.73$ & $  -0.0$ & $   3800 -   9235$ & $   8.0$ & F? & \nodata          & 1  \\
\multicolumn{8}{c}{{\bf SN 2005el}}\\
2005-09-26          & $  53639.82$ & $  -7.3$ & $   3800 -   9235$ & $   8.0$ & A  & $ 3.58\pm 0.26$  & 1  \\
2005-09-27          & $  53640.84$ & $  -6.3$ & $   3800 -   9235$ & $   8.0$ & A  & $ 3.44\pm 0.18$  & 1  \\
\multicolumn{8}{c}{{\bf SN 2005eq}}\\
2005-10-05          & $  53648.81$ & $  -5.2$ & $   3800 -   9235$ & $   8.0$ & N  & $< 1.59$         & 1  \\
\multicolumn{8}{c}{{\bf SN 2005hc}}\\
2005-10-18          & $  53661.71$ & $  -5.2$ & $   4000 -  10200$ & $   9.0$ & N? & $< 7.83$         & 1  \\
2005-10-18          & $  53661.71$ & $  -5.2$ & $   5800 -  10200$ & $   9.0$ & -- & \nodata          & 1  \\
\multicolumn{8}{c}{{\bf SN 2005ke}}\\
2005-11-23          & $  53697.72$ & $  -1.4$ & $   3780 -   7289$ & $   5.0$ & N  & \nodata          & 1  \\
2005-11-24$\dagger$ & $  53698.72$ & $  -0.4$ & $   3780 -   7290$ & $   5.0$ & N  & \nodata          & 1  \\
2005-11-25          & $  53699.61$ & $   0.4$ & $   3200 -   9589$ & $  14.0$ & N  & \nodata          & 1  \\
\multicolumn{8}{c}{{\bf SN 2005ki}}\\
2005-11-23          & $  53697.85$ & $  -7.6$ & $   3780 -   7289$ & $   5.0$ & F? & $< 2.40$         & 1  \\
2005-11-25$\dagger$ & $  53699.82$ & $  -5.7$ & $   3780 -   7289$ & $   5.0$ & N  & $< 0.72$         & 1  \\
\multicolumn{8}{c}{{\bf SN 2006D}}\\
2006-01-16          & $  53751.83$ & $  -6.1$ & $   3200 -   5300$ & $   6.0$ & -- & \nodata          & 1  \\
2006-01-16          & $  53751.84$ & $  -6.1$ & $   4000 -  10200$ & $   9.0$ & A  & $ 3.74\pm 0.21$  & 1  \\
2006-01-16          & $  53751.85$ & $  -6.0$ & $   5800 -  10200$ & $   9.0$ & A  & $ 3.34\pm 0.22$  & 1  \\
\multicolumn{8}{c}{{\bf SN 2006X}}\\
2006-02-09          & $  53775.71$ & $ -10.4$ & $   4135 -   6795$ & \nodata  & N  & \nodata          & 5  \\
2006-02-09          & $  53775.81$ & $ -10.3$ & $   3834 -   8139$ & \nodata  & N  & \nodata          & 5  \\
2006-02-11          & $  53777.65$ & $  -8.5$ & $   4175 -   6794$ & \nodata  & N  & \nodata          & 5  \\
2006-02-12          & $  53778.71$ & $  -7.4$ & $   4171 -   6794$ & \nodata  & N  & \nodata          & 5  \\
2006-02-13          & $  53779.63$ & $  -6.5$ & $   4178 -   6795$ & \nodata  & N  & \nodata          & 5  \\
2006-02-13          & $  53779.84$ & $  -6.3$ & $   3300 -   5300$ & $   6.0$ & N  & \nodata          & 1  \\
2006-02-13          & $  53779.86$ & $  -6.3$ & $   4000 -  10200$ & $   9.0$ & N  & $< 0.48$         & 1  \\
2006-02-13          & $  53779.86$ & $  -6.3$ & $   5800 -  10200$ & $   9.0$ & N  & $< 0.84$         & 1  \\
2006-02-18          & $  53784.59$ & $  -1.5$ & $   3902 -   8182$ & \nodata  & N  & \nodata          & 5  \\
2006-02-19          & $  53785.77$ & $  -0.4$ & $   4182 -   6800$ & \nodata  & N  & \nodata          & 5  \\
\multicolumn{8}{c}{{\bf SN 2006ax}}\\
2006-03-22$\dagger$ & $  53816.76$ & $ -10.5$ & $   3800 -   9235$ & $   8.0$ & F  & $ 1.21\pm 0.47$  & 1  \\
2006-03-23          & $  53817.65$ & $  -9.6$ & $   3800 -   9235$ & $   8.0$ & F  & $ 1.29\pm 0.26$  & 1  \\
2006-03-24          & $  53818.75$ & $  -8.5$ & $   3800 -   9235$ & $   8.0$ & F  & $ 1.91\pm 0.35$  & 1  \\
2006-03-30          & $  53824.78$ & $  -2.6$ & $   3800 -   9235$ & $   8.0$ & N  & $< 0.54$         & 1  \\
2006-03-31          & $  53825.73$ & $  -1.6$ & $   3800 -   9235$ & $   8.0$ & F? & \nodata          & 1  \\
2006-04-02          & $  53827.64$ & $   0.2$ & $   3800 -   9235$ & $   8.0$ & N  & \nodata          & 1  \\
\multicolumn{8}{c}{{\bf SN 2006dd}}\\
2006-06-21          & $  53907.92$ & $ -10.8$ & $   3413 -   9610$ & $   3.0$ & N  & \nodata          & 6  \\
\multicolumn{8}{c}{{\bf SN 2006gt}}\\
2006-09-25          & $  54003.75$ & $  -0.3$ & $   3800 -   9235$ & $   8.0$ & F? & \nodata          & 1  \\
\multicolumn{8}{c}{{\bf SN 2006hx}}\\
2006-10-05          & $  54013.65$ & $  -9.6$ & $   4000 -  10200$ & $   9.0$ & N  & $< 1.05$         & 1  \\
2006-10-10          & $  54018.77$ & $  -4.7$ & $   3800 -   9235$ & $   8.0$ & N  & $< 0.99$         & 1  \\
\multicolumn{8}{c}{{\bf SN 2006kf}}\\
2006-10-27          & $  54035.81$ & $  -5.1$ & $   3800 -   9235$ & $   8.0$ & N  & $< 1.26$         & 1  \\
\multicolumn{8}{c}{{\bf SN 2006mr}}\\
2006-11-09          & $  54048.65$ & $  -2.3$ & $   3785 -   6131$ & $   2.0$ & N  & \nodata          & 6  \\
2006-11-09          & $  54048.66$ & $  -2.3$ & $   5675 -   9976$ & $   4.0$ & N  & \nodata          & 6  \\
\multicolumn{8}{c}{{\bf SN 2006os}}\\
2006-11-22          & $  54061.68$ & $  -1.9$ & $   3800 -   9235$ & $   8.0$ & N  & \nodata          & 1  \\
\multicolumn{8}{c}{{\bf SN 2007A}}\\
2007-01-06          & $  54106.54$ & $  -6.2$ & $   3500 -   5300$ & $   6.0$ & -- & \nodata          & 1  \\
2007-01-06          & $  54106.56$ & $  -6.2$ & $   4000 -  10200$ & $   9.0$ & F  & $ 0.29\pm 0.53$  & 1  \\
2007-01-06          & $  54106.57$ & $  -6.2$ & $   5800 -  10200$ & $   9.0$ & F  & $ 1.85\pm 0.64$  & 1  \\
\multicolumn{8}{c}{{\bf SN 2007S}}\\
2007-01-31          & $  54131.82$ & $ -11.9$ & $   3500 -   5300$ & $   6.0$ & -- & \nodata          & 1  \\
2007-02-12          & $  54143.73$ & $  -0.1$ & $   3800 -   9235$ & $   8.0$ & N  & \nodata          & 1  \\
\multicolumn{8}{c}{{\bf SN 2007af}}\\
2007-03-04          & $  54163.82$ & $ -11.1$ & $   3300 -   5300$ & $   6.0$ & -- & \nodata          & 1  \\
2007-03-04          & $  54163.84$ & $ -11.1$ & $   4000 -  10200$ & $   9.0$ & A  & $ 1.33\pm 0.09$  & 1  \\
2007-03-04          & $  54163.86$ & $ -11.0$ & $   5800 -  10200$ & $   9.0$ & A  & $ 0.99\pm 0.11$  & 1  \\
2007-03-10          & $  54169.78$ & $  -5.2$ & $   3789 -  10889$ & $   6.0$ & A  & $ 0.50\pm 0.03$  & 1  \\
2007-03-14          & $  54173.80$ & $  -1.2$ & $   3429 -   9645$ & $   8.0$ & N  & \nodata          & 1  \\
2007-03-14          & $  54173.81$ & $  -1.1$ & $   4962 -   9645$ & $   8.0$ & N  & \nodata          & 1  \\
\multicolumn{8}{c}{{\bf SN 2007ai}}\\
2007-03-10          & $  54169.88$ & $  -2.7$ & $   4528 -  10187$ & $   6.0$ & N  & \nodata          & 1  \\
2007-03-12          & $  54171.89$ & $  -0.7$ & $   4528 -  10192$ & $   6.0$ & N  & \nodata          & 1  \\
\multicolumn{8}{c}{{\bf SN 2007as}}\\
2007-03-19          & $  54178.57$ & $  -2.2$ & $   3419 -   9626$ & $   8.0$ & N  & \nodata          & 1  \\
\multicolumn{8}{c}{{\bf SN 2007ax}}\\
2007-03-26$\dagger$ & $  54185.56$ & $  -2.1$ & $   3400 -   9608$ & $   8.0$ & F? & \nodata          & 1  \\
\multicolumn{8}{c}{{\bf SN 2007bd}}\\
2007-04-09          & $  54199.64$ & $  -7.5$ & $   3502 -   9697$ & $   4.0$ & N  & \nodata          & 1  \\
2007-04-12          & $  54202.55$ & $  -4.6$ & $   3800 -   9235$ & $   8.0$ & N  & $< 0.99$         & 1  \\
2007-04-17          & $  54207.59$ & $   0.2$ & $   3800 -   9235$ & $   8.0$ & N  & \nodata          & 1  \\
\multicolumn{8}{c}{{\bf SN 2007bc}}\\
2007-04-09          & $  54199.62$ & $  -0.5$ & $   3506 -   9694$ & $   4.0$ & N  & \nodata          & 1  \\
\multicolumn{8}{c}{{\bf SN 2007jg}}\\
2007-09-22          & $  54365.88$ & $  -0.2$ & $   3719 -   9961$ & $   5.0$ & N  & \nodata          & 1  \\
\multicolumn{8}{c}{{\bf SN 2007le}}\\
2007-10-16          & $  54389.65$ & $  -9.2$ & $   3412 -   9623$ & $   8.0$ & F? & $< 0.39$         & 1  \\
2007-10-21          & $  54394.58$ & $  -4.3$ & $   3352 -   9560$ & $   8.0$ & N  & $< 0.60$         & 1  \\
2007-10-21          & $  54394.60$ & $  -4.3$ & $   4704 -   9561$ & $   8.0$ & N  & $< 0.54$         & 1  \\
\multicolumn{8}{c}{{\bf SN 2007on}}\\
2007-11-11          & $  54415.79$ & $  -3.9$ & $   3800 -   9235$ & $   8.0$ & F? & $< 0.24$         & 1  \\
2007-11-14          & $  54418.84$ & $  -0.8$ & $   3800 -   9235$ & $   8.0$ & N  & \nodata          & 1  \\
\multicolumn{8}{c}{{\bf SN 2008ar}}\\
2008-02-29          & $  54525.87$ & $  -8.1$ & $   3743 -  10730$ & $   6.0$ & N  & \nodata          & 1  \\
\multicolumn{8}{c}{{\bf SN 2008bc}}\\
2008-03-14          & $  54539.52$ & $ -10.1$ & $   3300 -   6030$ & $  20.0$ & -- & \nodata          & 1  \\
2008-03-14          & $  54539.54$ & $ -10.1$ & $   5220 -   9240$ & $  30.0$ & N  & $< 0.60$         & 1  \\
2008-03-19          & $  54544.55$ & $  -5.1$ & $   3140 -   9470$ & $   0.5$ & N  & $< 0.06$         & 1  \\
2008-03-20          & $  54545.65$ & $  -4.0$ & $   3629 -   9437$ & $   7.0$ & N  & $< 0.27$         & 1  \\
\multicolumn{8}{c}{{\bf SN 2008bf}}\\
2008-03-20          & $  54545.73$ & $  -8.5$ & $   3629 -   9437$ & $   7.0$ & A  & $ 2.43\pm 0.14$  & 1  \\
2008-03-24          & $  54549.61$ & $  -4.8$ & $   3805 -  10100$ & $   7.0$ & A  & $ 1.48\pm 0.05$  & 1  \\
2008-03-25          & $  54550.71$ & $  -3.7$ & $   3805 -  10214$ & $   7.0$ & A  & $ 1.25\pm 0.07$  & 1  \\
2008-03-26          & $  54551.71$ & $  -2.7$ & $   3805 -  10214$ & $   7.0$ & A  & $ 0.78\pm 0.07$  & 1  \\
2008-03-28          & $  54553.75$ & $  -0.7$ & $   3400 -   5300$ & $   6.0$ & N  & \nodata          & 1  \\
2008-03-28          & $  54553.77$ & $  -0.7$ & $   4000 -  10200$ & $   9.0$ & N  & $< 0.75$         & 1  \\
\multicolumn{8}{c}{{\bf SN 2008bq}}\\
2008-04-07          & $  54563.48$ & $   0.3$ & $   3800 -   9235$ & $   8.0$ & N  & \nodata          & 1  \\
\multicolumn{8}{c}{{\bf SN 2008fp}}\\
2008-09-17          & $  54726.83$ & $  -3.7$ & $   3646 -   9460$ & $   7.0$ & A  & $ 0.93\pm 0.07$  & 1  \\
2008-09-21          & $  54730.85$ & $   0.4$ & $   3800 -   9235$ & $   8.0$ & N? & \nodata          & 1  \\
\multicolumn{8}{c}{{\bf SN 2008gl}}\\
2008-10-27          & $  54766.67$ & $  -1.5$ & $   3800 -   9235$ & $   8.0$ & N  & \nodata          & 1  \\
\multicolumn{8}{c}{{\bf SN 2008gp}}\\
2008-11-02          & $  54772.73$ & $  -6.0$ & $   4025 -  10713$ & $   6.0$ & N  & \nodata          & 1  \\
\multicolumn{8}{c}{{\bf SN 2008hj}}\\
2008-11-24          & $  54794.64$ & $  -6.0$ & $   3800 -   9235$ & $   8.0$ & N  & $< 0.75$         & 7  \\
2008-11-25          & $  54795.57$ & $  -5.0$ & $   3800 -   9235$ & $   8.0$ & N  & $< 0.63$         & 7  \\
\multicolumn{8}{c}{{\bf SN 2008hv}}\\
2008-12-10          & $  54810.81$ & $  -6.6$ & $   3300 -   6004$ & $  27.0$ & -- & \nodata          & 1  \\
2008-12-10          & $  54810.82$ & $  -6.6$ & $   5240 -   9210$ & $  39.0$ & A  & $ 0.93\pm 0.25$  & 1  \\
2008-12-15          & $  54815.85$ & $  -1.6$ & $   3620 -   9429$ & $   7.0$ & F? & $< 0.18$         & 1  \\
\multicolumn{8}{c}{{\bf SN 2008ia}}\\
2008-12-10          & $  54810.77$ & $  -2.8$ & $   3300 -   6004$ & $   2.7$ & -- & \nodata          & 1  \\
2008-12-10          & $  54810.79$ & $  -2.8$ & $   5240 -   9210$ & $   3.9$ & F  & \nodata          & 1  \\
\multicolumn{8}{c}{{\bf SN 2009D}}\\
2009-01-05          & $  54836.63$ & $  -3.9$ & $   3620 -   9428$ & $   7.0$ & N? & $< 0.24$         & 1  \\
2009-01-09          & $  54840.61$ & $  -0.1$ & $   5240 -   9210$ & $  39.0$ & N  & \nodata          & 1  \\
2009-01-09          & $  54840.64$ & $  -0.0$ & $   3300 -   6004$ & $  27.0$ & -- & \nodata          & 1  \\
\multicolumn{8}{c}{{\bf SN 2009F}}\\
2009-01-05          & $  54836.66$ & $  -5.4$ & $   3620 -   9429$ & $   7.0$ & A  & $ 3.58\pm 0.15$  & 1  \\
2009-01-09          & $  54840.71$ & $  -1.4$ & $   5240 -   9210$ & $  39.0$ & N  & \nodata          & 1  \\
2009-01-10          & $  54841.59$ & $  -0.5$ & $   3848 -   8017$ & $   9.0$ & N  & \nodata          & 1  \\
\multicolumn{8}{c}{{\bf SN 2009I}}\\
2009-01-17          & $  54848.56$ & $  -5.0$ & $   3629 -   9439$ & $   7.0$ & F  & $ 1.35\pm 0.26$  & 7  \\
2009-01-22          & $  54853.57$ & $  -0.0$ & $   3620 -   9428$ & $   7.0$ & N  & \nodata          & 7  \\
\multicolumn{8}{c}{{\bf SN 2009Y}}\\
2009-02-08          & $  54870.83$ & $  -5.5$ & $   3716 -   9439$ & $   7.0$ & N  & $< 0.24$         & 1  \\
2009-02-09          & $  54871.82$ & $  -4.5$ & $   3300 -   6004$ & $  27.0$ & -- & \nodata          & 1  \\
2009-02-09          & $  54871.83$ & $  -4.5$ & $   3716 -   9438$ & $   7.0$ & N  & $< 0.18$         & 1  \\
2009-02-09          & $  54871.84$ & $  -4.5$ & $   5240 -   9210$ & $  39.0$ & N  & $< 0.33$         & 1  \\
2009-02-09          & $  54871.84$ & $  -4.5$ & $   3694 -   6135$ & $   2.0$ & N  & \nodata          & 1  \\
2009-02-11          & $  54873.87$ & $  -2.5$ & $   4032 -  10118$ & $   5.0$ & N  & \nodata          & 1  \\
\multicolumn{8}{c}{{\bf SN 2009aa}}\\
2009-02-08          & $  54870.74$ & $  -7.8$ & $   3716 -   9439$ & $   7.0$ & F? & $< 0.18$         & 1  \\
2009-02-09          & $  54871.76$ & $  -6.8$ & $   3300 -   6004$ & $  27.0$ & -- & \nodata          & 1  \\
2009-02-09          & $  54871.78$ & $  -6.8$ & $   5240 -   9210$ & $  39.0$ & N  & $< 0.81$         & 1  \\
2009-02-11          & $  54873.76$ & $  -4.8$ & $   4032 -  10119$ & $   5.0$ & N  & \nodata          & 1  \\
2009-02-14          & $  54876.84$ & $  -1.8$ & $   4021 -  10119$ & $   5.0$ & N  & \nodata          & 1  \\
2009-02-16          & $  54878.84$ & $   0.1$ & $   4021 -  10123$ & $   5.0$ & N  & \nodata          & 1  \\
\multicolumn{8}{c}{{\bf SN 2009ab}}\\
2009-02-11          & $  54873.54$ & $ -10.2$ & $   4032 -  10123$ & $   5.0$ & N  & \nodata          & 1  \\
2009-02-14          & $  54876.63$ & $  -7.1$ & $   4021 -  10120$ & $   5.0$ & N  & \nodata          & 1  \\
2009-02-15          & $  54877.57$ & $  -6.2$ & $   4021 -  10125$ & $  11.0$ & N  & \nodata          & 1  \\
\multicolumn{8}{c}{{\bf SN 2009ad}}\\
2009-02-15          & $  54877.59$ & $  -8.3$ & $   4021 -  10124$ & $  11.0$ & N  & \nodata          & 1  \\
2009-02-16          & $  54878.63$ & $  -7.3$ & $   4021 -  10121$ & $   6.0$ & N  & \nodata          & 1  \\
2009-02-23          & $  54885.59$ & $  -0.5$ & $   3800 -   9235$ & $   8.0$ & N  & \nodata          & 1  \\
\multicolumn{8}{c}{{\bf SN 2009ag}}\\
2009-02-23          & $  54885.63$ & $  -3.5$ & $   3800 -   9235$ & $   8.0$ & N  & $< 0.36$         & 1  \\
2009-02-24          & $  54886.70$ & $  -2.4$ & $   3800 -   9235$ & $   8.0$ & N  & \nodata          & 1  \\
2009-02-26          & $  54888.68$ & $  -0.5$ & $   3800 -   9235$ & $   8.0$ & N  & \nodata          & 1  \\
\multicolumn{8}{c}{{\bf SN 2009cz}}\\
2009-04-17          & $  54938.52$ & $  -4.0$ & $   3714 -   9435$ & $   7.0$ & N  & $< 0.18$         & 7  \\
\multicolumn{8}{c}{{\bf SN 2009dc}}\\
2009-04-17          & $  54938.82$ & $  -7.7$ & $   3714 -   9433$ & $   7.0$ & SC & $19.66\pm 0.10$  & 7  \\
2009-04-18          & $  54939.79$ & $  -6.7$ & $   3450 -   9658$ & $   8.0$ & SC & $18.86\pm 0.30$  & 7  \\
2009-04-22          & $  54943.78$ & $  -2.8$ & $   3360 -   9568$ & $   8.0$ & SC & $17.17\pm 0.28$  & 7  \\
2009-04-23          & $  54944.80$ & $  -1.8$ & $   3361 -   9568$ & $   8.0$ & SC & $16.55\pm 0.29$  & 7  \\
\enddata
\tablecomments{Columns: (1) UT date of the observation; (2) Julian date of the observation (JD $-$ 2400000); (3) Phase in days since $B$-band maximum light; (4) Wavelength range covered; (5) Spectral resolution in \AA\ as estimated from the FWHM of arc-lamp lines; (6) Type of 6300 \AA\ feature (see Section~\ref{sec:CIIse}); (7) Pseudo-equivalent width of absorption at $\approx$6300 (3-$\sigma$ upper limits denoted with "$<$" sign); (8) References: 1 - \citet{folatelli11}; 2 - \citet{altavilla07}; 3 - \citet{pastorello07}; 4 - \citet{taubenberger08}; 5 - \citet{yamanaka09a}; 6 - \citet{stritzinger10}; 7 - This paper. \\ $\dagger$ Spectrum corrected to match photometry \citep[see][]{folatelli11}. }
\end{deluxetable}

\clearpage
\begin{deluxetable}{lcccc}
\tabletypesize{\scriptsize}
\tablecolumns{5}
\tablewidth{0pt}
\tablecaption{Summary of SYNOW parameters\label{tab:synow}}
\tablehead{
\colhead{Parameter} & \colhead{SN~2004eo} &\colhead{SN~2005el} & \colhead{SN~2006ax} & \colhead{SN~2009F} 
}
\startdata
Age [d] & $-6$ & $-7$ & $-10$ & $-5$ \\
\ion{C}{2} class & N & A & F & A \\
$v_{\mathrm{ph}}$ [km s$^{-1}$] & 11,500 & 12,300 & 12,500 & 12,000 \\
$T_{\mathrm{BB}}$ [K] & 12,500 & 11,500 & 12,500 & 8,500 \\
$\tau$(\ion{C}{2})     & \nodata & \nodata & 0.04    & 0.1     \\
$\tau$(\ion{C}{2} HV)  & \nodata  & 0.05   & \nodata & \nodata \\
$v_{\mathrm{min}}$(\ion{C}{2} HV) [km s$^{-1}$] & \nodata & 12,800 & \nodata &  \\
$\tau$(\ion{C}{3})     & \nodata & \nodata & 0.5     & \nodata \\
$\tau$(\ion{O}{1})     & 0.7     & 0.5     & 0.1     & 1.5     \\
$\tau$(\ion{Na}{1})    & 0.2     & 0.1     & 0.1     & 0.5     \\
$\tau$(\ion{Mg}{2})    & 1.0     & 0.8     & 0.7     & 0.8     \\
$\tau$(\ion{Si}{2})    & 4.3     & 2.5     & 3.0     & 5.0     \\
$\tau$(\ion{Si}{3})    & 0.6     & 0.5     & 0.7     & 0.8     \\
$\tau$(\ion{S}{2})     & 1.5     & 1.0     & 0.8     & 0.5     \\
$\tau$(\ion{Ca}{2})    & 14.0    & 4.5     & 7.0     & 20.0    \\
$\tau$(\ion{Ca}{2} HV) & 2.0     & 1.0     & 2.5     & \nodata \\
$v_{\mathrm{min}}$(\ion{Ca}{2} HV) [km s$^{-1}$] & 20,000 & 20,000 & 20,000 & \nodata \\
$\tau$(\ion{Ti}{2})    & 0.3     & \nodata & \nodata & 1.0     \\
$\tau$(\ion{Fe}{2})    & 0.3     & 0.1     & 0.1     & 0.1     \\
$\tau$(\ion{Fe}{2} HV) & 0.4     & 0.2     & 0.3     & \nodata \\
$v_{\mathrm{min}}$(\ion{Fe}{2} HV) [km s$^{-1}$] & 18,000 & 19,000 & 19,000 & \nodata \\
$\tau$(\ion{Fe}{3})    & 1.0     & 0.5     & 0.6     & 0.5     \\
\enddata
\end{deluxetable}

\begin{deluxetable}{lcccc}
\tabletypesize{\small}
\tablecolumns{5}
\tablewidth{0pt}
\tablecaption{Fits of intrinsic $(B-V)$ pseudo-color versus decline rate \label{tab:bvdm}}
\tablehead{
\colhead{Sample} & \colhead{$a$} &\colhead{$b$} & \colhead{$\sigma_I$} & \colhead{$N_{\mathrm{SNe}}$} \\ 
\colhead{(1)} & \colhead{(2)} & \colhead{(3)} & \colhead{(4)} & \colhead{(5)}  
}
\startdata
With carbon     & $-0.023 \pm 0.028$    & \phs$0.31 \pm 0.15$ & $0.06 \pm 0.05$ & 6  \\
Without carbon  & \phs$0.020 \pm 0.018$ & $-0.01 \pm 0.07$    & $0.03 \pm 0.04$ & 6  \\
All             & $-0.002 \pm 0.018$    & \phs$0.09 \pm 0.08$ & $0.05 \pm 0.02$ & 12 \\
\citet{folatelli10b} & $-0.02 \pm 0.01$ & \phs$0.12 \pm 0.05$ & \nodata         & 14 \\
\enddata
\tablecomments{Fits of the type: $(B-V)_0\;=\;a\;+\;b\,[$\dm\,$-\,1.1]$ for \dm~$\leq 1.6$ mag. \\ Columns: (1) sample of SNe; (2) fit intercept; (3) fit slope; (4) intrinsic dispersion of fit (in magnitudes; see Section~\ref{sec:phprop}); (5) number of SNe used in fit.}
\end{deluxetable}


\begin{thebibliography}{}


\bibitem[Altavilla et al.(2007)]{altavilla07}
  Altavilla,~G., et~al. 2007, \aap, 475, 585
\bibitem[Arnett, Truran \& Woosley(1971)]{arnett71} Arnett,~W.~D.,
  Truran,~J.~W. \& Woosley,~S.~E. 1971, \apj, 165, 87
\bibitem[Benetti et al.(2005)]{benetti05} Benetti,~S., et~al. 2005,
  \apj, 623, 1011
\bibitem[Branch et al.(2002)]{branch02} Branch,~D., et~al. 2002, \apj,
  566, 1005 
\bibitem[Branch et al.(2003)]{branch03} Branch,~D., et~al. 2003, \aj,
  126, 1489 
\bibitem[Branch et al.(2006)]{branch06} Branch,~D., et~al. 2006,
  \pasp, 118, 560 
\bibitem[Branch et al.(2007)]{branch07} Branch,~D., et~al. 2007,
  \pasp, 119, 709 
\bibitem[Burns et al.(2011)]{burns11} Burns, C.~R., et~al. 2011, \aj, 
  141, 19
\bibitem[Contreras et al.(2010)]{contreras10} Contreras,~C.,
  et~al. 2010, \aj, 139, 519
\bibitem[Folatelli(2010a)]{folatelli10a} Folatelli, G. 2010, Proceedings of
the XXVIth IAP Annual Colloquium, ``Progenitors and environments of stellar
explosions'' (\url{http://www.iap.fr/col2010/Proceedings/posters/Folatelli.pdf})
\bibitem[Folatelli et al.(2010b)]{folatelli10b} Folatelli,~G.,
  et~al. 2010, \aj, 139, 120
\bibitem[Folatelli et al.(2011)]{folatelli11} Folatelli,~G.,
  et~al. 2011, in preparation
\bibitem[Foley \& Kasen(2011)]{foley11} Foley,~R.~J. \& Kasen,~D. 2011,
  \apj, 729, 55
\bibitem[Gamezo, Khokhlov \& Oran(2005)]{gamezo05} Gamezo,~V.~N.,
  Khokhlov,~A.~M. \& Oran,~E.~S. 2005, \apj, 623, 337 
\bibitem[Garavini et al.(2004)]{garavini04} Garavini,~G.,
  et~al. 2004, \aj, 128, 387
\bibitem[Hamuy et al.(2006)]{hamuy06} Hamuy, M., et~al. 2006, \pasp,
  118, 2
\bibitem[Hatano et al.(1999)]{hatano99} Hatano,~K., et~al 1999, \apjs,
  121, 233
\bibitem[Hicken et al.(2007)]{hicken07} Hicken,~M., et~al. 2007, \apjl,
  669, 17
\bibitem[Hillebrandt \& Niemeyer(2000)]{hillebrandt00}
  Hillebrandt,~W. \& Niemeyer,~J.~C. 2000, \araa, 38, 191
\bibitem[H\"oflich, Khokhlov \& Wheeler(1995)]{hoeflich95} H\"oflich,~P.,
  Khokhlov,~A.~M. \& Wheeler,~J.~C. 1995, \apj, 444, 831
\bibitem[H\"oflich \& Khokhlov(1996)]{hoeflich96} H\"oflich,~P.
  \& Khokhlov,~A.~M. 1996, \apj, 457, 500
\bibitem[H\"oflich et al.(2010)]{hoeflich10} H\"oflich,~P.,
  et~al. 2010, \apj, 710, 444 
\bibitem[Howell et al.(2006)]{howell06} Howell,~D.~A., et~al. 2006,
  \nat, 443, 308
\bibitem[Hoyle \& Fowler(1960)]{hoyle60} Hoyle,~F. \&
  Fowler,~W.~A. 1960, \apj, 132, 565
\bibitem[Iben \& Tutukov(1984)]{iben84} Iben,~I.,~Jr. \&
  Tutukov,~A.~V. 1984, \apjs, 54, 335 
\bibitem[Iwamoto et al.(1999)]{iwamoto99} Iwamoto,~K., et~al. 1999,
  \apjs, 125, 439
\bibitem[Kasen \& Woosley(2007)]{kasen07} Kasen,~D. \&
  Woosley,~S.~E. 2007, \apj, 656, 661
\bibitem[Kasen et al.(2009)]{kasen09} Kasen,~D., R\"opke,~F.~K. \&
  Woosley,~S.~E. 2009, \nat, 460, 869
\bibitem[Khokhlov(1991)]{khokhlov91} Khokhlov,~A.~M. 1991, \aap, 245,
  114
\bibitem[Lentz et al.(2002)]{lentz02} Lentz,~E.~J., Baron,~E.,
  Hauschildt,~P.~H. \& Branch,~D. 2002, \apj, 580, 374
\bibitem[Lucy(1999)]{lucy99} Lucy,~L.~B. 1999, \aap, 345, 211
\bibitem[Maeda et al.(2010b)]{maeda10b} Maeda,~K., et~al. 2010b, \apj,
  712, 624
\bibitem[Maeda et al.(2010c)]{maeda10c} Maeda,~K., et~al. 2010c, \nat,
  466, 82
\bibitem[Maeda et al.(2011)]{maeda11} Maeda,~K., et~al. 2011, \mnras,
  413, 3075
\bibitem[Marietta et al.(2000)]{marietta00} Marietta,~E.,
  Burrows,~A. \& Fryxell,~B. 2000, \apjs, 128, 615
\bibitem[Marion et al.(2006)]{marion06} Marion,~G.~H., et~al. 2006,
  \apj, 645, 1392
\bibitem[Marion et al.(2009)]{marion09} Marion,~G.~H., et~al. 2009,
  \aj, 138, 727
\bibitem[Maund et al.(2010)]{maund10} Maund,~J.~R., et~al. 2010,
  \apjl, 725, 167
\bibitem[Mazzali \& Lucy(1993)]{mazzali93} Mazzali,~P.~A. \&
  Lucy,~L.~B. 1993, \aap, 279, 447
\bibitem[Mazzali(2000)]{mazzali00} Mazzali,~P.~A. 2000, \aap, 363, 705
\bibitem[Mazzali(2001)]{mazzali01} Mazzali,~P.~A. 2001, \mnras, 321,
  341 
\bibitem[Mazzali et al.(2005)]{mazzali05} Mazzali,~P.~A., et~al. 2005,
  \apjl, 623, 37
\bibitem[Mazzali et al.(2007)]{mazzali07} Mazzali,~P.~A.,
  R\"opke,~F.~K., Benetti,~S. \& Hillebrandt,~W. 2007, Science, 315,
  825
\bibitem[Mazzali et al.(2008)]{mazzali08} Mazzali,~P.~A.,
et~al. 2008, \mnras, 386, 1897
\bibitem[Nomoto, Sugimoto \& Neo(1976)]{nomoto76} Nomoto,~K.,
  Sugimoto,~D. \& Neo,~S. 1976, \apss, 39, L37 
\bibitem[Nomoto(1982)]{nomoto82} Nomoto,~K. 1982, \apj, 253, 798
\bibitem[Nomoto et al.(1984)]{nomoto84} Nomoto,~K.,
  Thielemann,~F.-K. \& Yokoi,~K. 1984, \apj, 286, 644
\bibitem[Nomoto et al.(2007)]{nomoto07} Nomoto,~K., Saio,~H.,
  Kato,~M. \& Hachisu,~I. 2007, \apj, 663, 1269
\bibitem[Nugent et al.(1995)]{nugent95} Nugent,~P., et~al. 1995, \apjl,
  455, 147 
\bibitem[Pakmor et al.(2008)]{pakmor08} Pakmor,~R., et~al. 2008, \aap, 489,
  943. 
\bibitem[Parrent et al.(2011)]{parrent11} Parrent,~J.~T., et~al. 2011,
  \apj, 732, 30 
\bibitem[Pastorello et al.(2007)]{pastorello07} Pastorello,~A.,
  et~al. 2007, \mnras, 377, 1531
\bibitem[Perlmutter et al.(1999)]{perlmutter99} Perlmutter,~S.,
  et~al. 1999, \apj, 517, 565 
\bibitem[Phillips(1993)]{phillips93} Phillips,~M.~M. 1993, \apjl, 413,
  105 
\bibitem[Pignata et al.(2008)]{pignata08} Pignata,~G., et~al. 2008,
  \mnras, 388, 971
\bibitem[Pinto \& Eastman(2000)]{pinto00} Pinto,~P.~A. \&
  Eastman,~R.~G. 2000, \apj, 530, 757
\bibitem[Riess et al.(1998)]{riess98} Riess,~A.~G., et~al. 1998, \aj,
  116, 1009
\bibitem[R\"opke(2005)]{roepke05} R\"opke,~F.~K. 2005, \aap, 432, 969
\bibitem[Scalzo et al.(2010)]{scalzo10} Scalzo,~R.~A., et~al. 2010,
  \apj, 713, 1073
\bibitem[Shigeyama et al.(1992)]{shigeyama92} Shigeyama,~T., Nomoto,~K.,
  Yamaoka,~H. \& Thielemann,~F.-K. 1992, \apjl, 386, 13
\bibitem[Silverman et al.(2011)]{silverman11} Silverman,~J.~M.,
  et~al. 2011, \mnras, 410, 585
\bibitem[Stanishev et al.(2007)]{stanishev07} Stanishev,~V.,
  et~al. 2007, \aap, 469, 645
\bibitem[Stritzinger et al.(2006)]{stritzinger06}
  Stritzinger,~M., Mazzali,~P.~A., Sollerman,~J. \& Benetti,~S. 2006,
  \aap, 460, 793
\bibitem[Stritzinger et al.(2010)]{stritzinger10}
  Stritzinger,~M. et~al. 2010, \aj, 140, 2036
\bibitem[Stritzinger et al.(2011)]{stritzinger11}
  Stritzinger,~M. et~al. 2011, \aj, 142, 156
\bibitem[Tanaka et al.(2006)]{tanaka06} Tanaka,~M., Mazzali,~P.~A.,
  Maeda,~K. \& Nomoto,~K. 2006, \apj, 645, 470
\bibitem[Tanaka et al.(2008)]{tanaka08} Tanaka,~M., et~al. 2008, \apj,
  677, 448 
\bibitem[Tanaka et al.(2009)]{tanaka09} Tanaka,~M., et~al. 2009, \apj,
  692, 1131
\bibitem[Tanaka et al.(2010)]{tanaka10} Tanaka,~M., et~al. 2010, \apj,
  714, 1209 
\bibitem[Tanaka et al.(2011)]{tanaka11} Tanaka,~M., et~al. 2011, \mnras,
  410, 1725 
\bibitem[Taubenberger et al.(2008)]{taubenberger08}
  Taubenberger,~S. et~al. 2008, \mnras, 385, 75 
\bibitem[Taubenberger et al.(2011)]{taubenberger11}
  Taubenberger,~S. et~al. 2011, \mnras, 412, 2735
\bibitem[Thomas et al.(2007)]{thomas07} Thomas,~R.~C. et~al. 2007,
  \apjl, 654, 53 
\bibitem[Wang et al.(2009)]{wang09} Wang,~X., et~al. 2009, \apjl, 699,
  139
\bibitem[Webbink(1984)]{webbink84} Webbink,~R.~F. 1984, \apj, 277, 355
\bibitem[Whelan \& Iben(1983)]{whelan73} Whelan,~J. \&
  Iben,~I.,~Jr. 1973, \apj, 186, 1007 
\bibitem[Yamanaka et al.(2009a)]{yamanaka09a} Yamanaka,~M., et~al. 2009a,
  \pasj, 61, 713
\bibitem[Yamanaka et al.(2009b)]{yamanaka09b} Yamanaka,~M.,
  et~al. 2009b, \apjl, 707, 118
\end{thebibliography}
\end{document}